\documentclass{article} 
\usepackage{iclr2024_conference,times}

\usepackage[colorlinks,
            anchorcolor=red,
            citecolor=citecolor, 
            linkcolor=linkcolor,
            ]{hyperref}
\usepackage{url}

\usepackage{booktabs}      
\usepackage{amsfonts}       
\usepackage{nicefrac}      
\usepackage{microtype}                     
\usepackage{url}                         
\usepackage{multirow}
\usepackage{color}
\usepackage{colortbl}
\usepackage{caption}
\usepackage{graphicx}
\usepackage{amsmath}
\usepackage{wrapfig}
\usepackage{lipsum}
\usepackage{algorithm}
\usepackage{algorithmic}
\usepackage{listings}
\usepackage{listing}
\usepackage{xcolor}
\usepackage{framed}
\usepackage{enumitem}
\usepackage{amssymb}
\usepackage{setspace}
\usepackage{epsfig}
\usepackage{subfigure}
\usepackage{mathtools}
\usepackage{verbatim}
\usepackage{booktabs} 
\usepackage{array}
\usepackage{footnote}
\usepackage{tabularx}
\usepackage{booktabs}
\usepackage{todonotes} 
\usepackage{xspace}
\usepackage{colortbl}
\usepackage{bbm}

\definecolor{citecolor}{HTML}{0071BC}
\definecolor{linkcolor}{HTML}{ED1C24}

\usepackage{minitoc}
\usepackage[toc,page,header]{appendix}
\usepackage{soul}
\usepackage{makecell}
\usepackage{framed,multirow}
\usepackage{threeparttable}
\usepackage{amssymb}
\usepackage{pifont}
\usepackage{algorithm}
\usepackage{layouts}
\usepackage{listings}
\usepackage{colortbl}
\usepackage{pythonhighlight}
\usepackage{tikz}
\usepackage{enumitem}

\definecolor{iblue}{rgb}{0.06, 0.75, 1.0}
\definecolor{igray}{rgb}{0.00, 0.00, 0.00}

\newcolumntype{P}[1]{>{\centering\arraybackslash}p{#1}}
\newlength\savewidth

\newcolumntype{R}[1]{>{\raggedleft\arraybackslash}p{#1}}
\newcolumntype{L}[1]{>{\raggedright\arraybackslash}p{#1}}

\newcommand{\ourmodel}{{\fontfamily{ppl}\selectfont SuPreM}}
\newcommand{\ourdataset}{{\fontfamily{ppl}\selectfont
AbdomenAtlas 1.1}}

\newcommand{\totalct}{{\fontfamily{ppl}\selectfont
9,262}}
\newcommand{\totalslices}{{\fontfamily{ppl}\selectfont
2,789,975}}
\newcommand{\totalmasks}{{\fontfamily{ppl}\selectfont
251,323}}

\newsavebox{\encoder}
\sbox{\encoder}{
    \begin{tikzpicture}[scale=0.2] 
        \draw [rotate=270] (0,0) -- (1,0) -- (0.8,1) -- (0.2,1) -- cycle;
    \end{tikzpicture}
}
\newsavebox{\decoder}
\sbox{\decoder}{
    \begin{tikzpicture}[scale=0.2] 
        \draw [rotate=90] (0,0) -- (1,0) -- (0.8,1) -- (0.2,1) -- cycle;
    \end{tikzpicture}
}
\title{How Well Do Supervised 3D Models Transfer to Medical Imaging Tasks?}

\author{
\begin{tabular}[h]{l}
    Wenxuan Li \quad
    Alan Yuille \quad
    Zongwei Zhou\thanks{Correspondence to Zongwei Zhou (\href{mailto:zzhou82@jh.edu}{\textsc{zzhou82@jh.edu}}).
    }
\end{tabular}
\\
\begin{tabular}[h]{l}
    ~Johns Hopkins University
\end{tabular}
\\
\begin{tabular}[h]{l}
    ~{\small \href{https://github.com/MrGiovanni/SuPreM}{https://github.com/MrGiovanni/SuPreM}}
\end{tabular}
}

\definecolor{ablation_red}{RGB}{196,64,60}
\definecolor{ablation_green}{RGB}{0,155,85}

\definecolor{mygray}{gray}{0.6}
\definecolor{mygray-bg}{gray}{0.9}

\iclrfinalcopy 
\begin{document}

\maketitle

\doparttoc 
\faketableofcontents 

\begin{abstract}
    The pre-training and fine-tuning paradigm has become prominent in transfer learning. For example, if the model is pre-trained on ImageNet and then fine-tuned to PASCAL, it can significantly outperform that trained on PASCAL from scratch. While ImageNet pre-training has shown enormous success, it is formed in 2D, and the learned features are for classification tasks; when transferring to more diverse tasks, like 3D image segmentation, its performance is inevitably compromised due to the deviation from the original ImageNet context. A significant challenge lies in the lack of large, annotated 3D datasets rivaling the scale of ImageNet for model pre-training. To overcome this challenge, we make two contributions. Firstly, we construct \ourdataset\ that comprises \totalct\ three-dimensional computed tomography (CT) volumes with high-quality, per-voxel annotations of 25 anatomical structures and pseudo annotations of seven tumor types. Secondly, we develop a suite of models that are pre-trained on our \ourdataset\ for transfer learning. Our preliminary analyses indicate that the model trained only with 21 CT volumes, 672 masks, and 40 GPU hours has a transfer learning ability similar to the model trained with 5,050 (unlabeled) CT volumes and 1,152 GPU hours. More importantly, the transfer learning ability of supervised models can further scale up with larger annotated datasets, achieving significantly better performance than preexisting pre-trained models, irrespective of their pre-training methodologies or data sources. We hope this study can facilitate collective efforts in constructing larger 3D medical datasets and more releases of supervised pre-trained models.
\end{abstract}

\section{Introduction}\label{sec:introduction}

Pre-training and fine-tuning is a widely adopted transfer learning paradigm \citep{zoph2020rethinking}. Given the relationship across different vision tasks, a model pre-trained on one dataset is expected to benefit another. Over the past few decades, pre-training has been important in AI development \citep{kumar2017weight,radford2021learning}. For 2D vision tasks, there are two available options: (\textit{i}) supervised pre-training and (\textit{ii}) self-supervised pre-training, but for 3D vision tasks, option (\textit{i}) is often not available simply due to the lack of large, annotated 3D volumetric datasets \citep{wang2022p2p}.

Supervised pre-training can learn image features that are transferable to many target tasks. It has been common practice to pre-train models using ImageNet and then fine-tune the model on target tasks that often have less training data, e.g., PASCAL.
However, two challenges arise in ImageNet pre-training. Firstly, ImageNet predominantly comprises 2D images, leaving a palpable void in large-scale 3D datasets and investigation in 3D transfer learning \citep{huang2023stu}. Secondly, ImageNet is intended for image classification, so the benefit for segmentation (and other vision tasks) can be somewhat compromised \citep{he2019rethinking}.
If such an ImageNet-like dataset exists---formed in 3D and annotated per voxel---supervised pre-trained models are expected to transfer better to 3D image segmentation than self-supervised ones for two reasons.

\begin{enumerate}[leftmargin=*]
    
    \item \textbf{Supervised pre-training is more efficient in data and computation because of its explicit learning objective.} While self-supervised pre-training can learn features without manual annotation, it often requires a large corpus of datasets \citep{xiao2022delving}. Extracting meaningful features directly from raw, unlabeled data is inherently challenging. Unlabeled data have a high degree of redundancy \citep{haghighi2020learning,haghighi2021transferable} and noise \citep{mahajan2018exploring}, which can complicate the learning process. Therefore, self-supervised pre-training often calls for greater computational resources and time to match the outcomes achieved by supervised pre-training \citep{chen2020simple,tang2022self}. We have quantified the improved data and computational efficiency from perspectives of both pre-training (\figureautorefname~\ref{fig:data_compute_effciency}a; 99.6\% fewer data) and fine-tuning (\figureautorefname~\ref{fig:data_compute_effciency}b; 66\% less computation). Specifically, the model trained with 21 CT volumes, 672 masks, and 40 GPU hours shows transfer learning ability similar to that trained with 5,050 CT volumes and 1,152 GPU hours, highlighting the remarkable efficiency of supervised pre-training. 

    \item \textbf{Supervised pre-training enables the model to learn image features that are relevant to image segmentation.} Self-supervised pre-training must extract images features from raw, unlabeled data using pretext tasks such as mask image modeling \citep{chen2019self,tao2020revisiting,zhou2021models,he2022masked}, instance discrimination \citep{xie2020pgl,chaitanya2020contrastive,shekoofeh2021big}, etc. Despite their efficacy in pre-training, these pretext tasks share no obvious relation to the target image segmentation. In contrast, supervised pre-training uses semantically meaningful annotations (e.g., organ/tumor segmentation) as supervision, with which the model can mimic the behavior of medical professionals---identifying the edge and boundary of specific anatomical structures. As a result, the pre-training is interpretable, and the learned features are expected to be relevant to image segmentation tasks \citep{zamir2018taskonomy,ilharco2022editing,you2022class}. We have demonstrated that the learned features can be \textit{direct inference} for organ segmentation on CT volumes collected from hospitals worldwide (\tableautorefname~\ref{tab:external_datasets}; evaluated on three novel hospitals). The features learned by supervision can also be \textit{fine-tuned} to perform novel class segmentation (unseen in the pre-training) with higher accuracy and less annotated data than the features learned by self-supervision (\tableautorefname~\ref{tab:novel_classes}; evaluated on 63 novel classes).
    
\end{enumerate}

This paper seeks to answer the question \textit{how well the model transfers to 3D medical imaging tasks} IF it is pre-trained on large, annotated 3D datasets. Naturally, we start with creating an \textit{IF} dataset at a massive scale. \textbf{Firstly}, we construct a dataset (termed \ourdataset\footnote{Segmentation is fundamental in the medical domain \citep{ma2023towards}. It can be viewed as a per-voxel classification task. Therefore, the per-voxel supervision used in our pre-training (\textbf{272.7B} annotated voxels) is much stronger than the per-image supervision used in ImageNet pre-training (\textbf{14M} images).}) of \totalct\ CT volumes with per-voxel annotations of 25 anatomical structures and pseudo annotations of seven types of tumors. This large-scale, fully-annotated dataset enables us to train models in a fully supervised manner using multi-organ segmentation as the pretext task. As reviewed in~\tableautorefname~\ref{tab:dataset_overview}, this dataset is much more extensive (considering both the number of CT volumes and annotated classes) than public datasets \citep{wasserthal2022totalsegmentator,ma2022fast,qu2023annotating}. Scaling experiments in \S\ref{sec:extensive_dataset} suggested that pre-training models on more annotated datasets can further improve the transfer learning ability.
\textbf{Secondly}, we develop a suite of \ul{\textbf{Su}}pervised \ul{\textbf{Pre}}-trained \ul{\textbf{M}}odels, termed \ourmodel, that combined the good of large-scale datasets and per-voxel annotations, demonstrating the efficacy across a range of target segmentation tasks. As reported in \S\ref{sec:model_zoo}, some of the dominant segmentation backbones have been pre-trained and will be available to the public. Current pre-trained backbones are U-Net (CNN-type) \citep{ronneberger2015u}, SegResNet (CNN-type) \citep{myronenko20193d}, and Swin UNETR (Transformer-type) \citep{tang2022self}, and more backbones will be added along time. 

In prospective endeavors, we anticipate that the expansion of datasets and annotations will not only enhance feature learning, as demonstrated in this study, but also promote the development of advanced AI algorithms and benchmark the state of the art in terms of segmentation performance, inference efficiency, and domain generalizability.

\section{Brief History: Supervised Pre-Training}\label{sec:history}

In a major initiative aimed at developing widely transferable AI models---known as Foundation Models in the medical domain \citep{moor2023foundation,butoi2023universeg,ma2023segment}---one faces a critical decision: \textit{should the focus of pre-training be supervised or self-supervised?} While human annotations undeniably improve task-specific performance, such as semantic segmentation, the best strategy for learning generic image features that can be transferable across a spectrum of tasks has yet to be determined. For 2D vision tasks, the advent of ImageNet \citep{deng2009imagenet} makes it possible to debate the merits and limitations of supervised pre-trained models for transfer learning compared with self-supervised ones. We refer the readers to \citet{yang2020transfer} and \citet{tendle2021study} for a plethora of viewpoints from either side. In essence, the debates are about clarifying the learning objective (loss function) of emulating human vision \citep{zhou2021towards}.

The learning objective of supervised pre-training is to minimize the discrepancy between AI predictions and semantic labels annotated by humans. Over the years, supervised pre-training on ImageNet has shown marked success in transfer learning \citep{yosinski2014transferable}. Moreover, the transfer learning ability can be further enhanced when models are trained on increasingly expansive datasets, such as ImageNet-21K \citep{kolesnikov2020big}, Instagram \citep{mahajan2018exploring}, JFT-300M \citep{sun2017revisiting}, and JFT-3B \citep{zhai2022scaling}. In general, supervised pre-training exhibits clear advantages over self-supervised pre-training when sizable annotated datasets are available \citep{steiner2021train,ridnik2021imagenet}. However, acquiring millions of manual annotations is labor-intensive, time-consuming, and challenging to scale---but certainly not impossible---evidenced by several recent influential endeavors \citep{kuznetsova2020open,mei2022radimagenet,kirillov2023segment,bai2023sequential}.

On the other hand, self-supervised pre-training offers an alternative by enabling AI models to learn from raw, unlabeled data \citep{jing2020self,zoph2020rethinking,ren2022simple,ren2023tinymim}, thus reducing the need for manual annotation. Self-supervised pre-training has historically lagged behind the state-of-the-art supervised pre-training in ImageNet benchmarks \citep{pathak2016context,noroozi2016unsupervised}. The recent pace of progress in self-supervised pre-training has yielded models whose performance not only matches but, at times, surpasses those achieved by supervised pre-training \citep{chen2020simple,grill2020bootstrap,chen2020big,zhou2021ibot,wei2022masked}. This has raised hopes that self-supervised pre-training could indeed replace the ubiquitous supervised pre-training in advanced computer vision going forward. The caveat, however, is the significant demand for both data and computational power, often exceeding the resources available in academic settings. For example, \citet{he2020momentum} have demonstrated that self-supervised features trained on 1B images (a factor of 714$\times$ larger) can transfer comparably or better than ImageNet features.

Supervised pre-training on ImageNet has demonstrated benefit for 2D medical image tasks after transfer learning \citep{tajbakhsh2016convolutional,shin2016deep,zhou2017fine}. Unfortunately, it has been constrained for 3D medical imaging tasks due to the lack of a 3D counterpart to ImageNet. Although there are a great number of raw, unlabeled medical images available \citep{national2011national,baxter2023scottish,zhao2023one,saenz2024maida}, annotating these images is a labor-intensive undertaking for professionals. Our contribution to a large, annotated 3D dataset could spark the debate of whether self-supervised or supervised pre-training leads to better performance and data/computational efficiency, which would not be possible without the invention of a dataset of such a scale.

\section{Material \& Method}\label{sec:material_method}

We constructed an \ourdataset\ dataset comprising \textbf{\totalct} three-dimensional CT volumes and over \textbf{\totalmasks} masks spanning \textbf{25} anatomical structures and \textbf{7} types of tumors. In addition, we released a suite of supervised pre-trained models (\ourmodel) to benefit 3D medical imaging tasks.

\subsection{Extensive Dataset: \ourdataset}\label{sec:extensive_dataset}

\begin{table*}[h]
    \caption{\textbf{Contribution \#1: An extensive dataset of \totalct\ CT volumes with per-voxel annotations of 25 anatomical structures.} 
    This dataset is unprecedented in terms of data and annotation scales, providing over \totalmasks\ organ/tumor masks and \totalslices\ annotated images that are taken from 88 hospitals worldwide. In 2009, before the advent of ImageNet \citep{deng2009imagenet}, it was challenging to empower an AI model with generalized image representation using a small or even medium size of labeled data, the same situation, we believe, that presents in 3D medical image analysis today. As seen in the table, the annotations of public datasets are limited, partial, and incomplete, and the CT volumes in these datasets are often biased toward specific populations, medical centers, and countries. Our constructed dataset mitigates these gaps, representing a significant leap forward in the field. The CT volumes in datasets 1--17 are used to construct \ourdataset. The domain gap across these datasets is illustrated in \appendixname\ \ref{sec:domain_gaps_appendix}.
    }
    \centering
    \scriptsize
    \begin{tabular}{p{0.248\linewidth}P{0.041\linewidth}P{0.06\linewidth}P{0.05\linewidth}|p{0.205\linewidth}P{0.041\linewidth}P{0.06\linewidth}P{0.05\linewidth}}
    \toprule
    dataset (year) [source] & \makecell{\# of\\organ} & \makecell{\# of$^{\dagger}$\\volume} & \makecell{\# of\\center} & dataset (year) [source] & \makecell{\# of\\organ} & \makecell{\# of$^{\dagger}$\\volume} & \makecell{\# of\\center} \\
    \midrule
    1. Pancreas-CT \citeyearpar{roth2015deeporgan} [\href{https://academictorrents.com/details/80ecfefcabede760cdbdf63e38986501f7becd49}{link}] & 1 & 42 & 1 & 2. CHAOS \citeyearpar{valindria2018multi} [\href{https://chaos.grand-challenge.org/Download/}{link}] & 4 & 20 & 1\\
    3. CT-ORG \citeyearpar{rister2020ct} [\href{https://wiki.cancerimagingarchive.net/pages/viewpage.action?pageId=61080890#61080890cd4d3499fa294f489bf1ea261184fd24}{link}] & 5 & 140 & 8 & 4. BTCV \citeyearpar{landman2015miccai} [\href{https://www.synapse.org/#!Synapse:syn3193805/wiki/89480}{link}] & 12 & 47 & 1 \\
    5. AMOS22 \citeyearpar{ji2022amos} [\href{https://amos22.grand-challenge.org}{link}] & 15 & 200 & 2 & 6. WORD \citeyearpar{luo2021word} [\href{https://github.com/HiLab-git/WORD}{link}] & 16 & 120 & 1 \\
    7-12. MSD CT Tasks \citeyearpar{antonelli2021medical} [\href{https://decathlon-10.grand-challenge.org/}{link}] & 9 & 945 & 1 & 13. LiTS \citeyearpar{bilic2019liver} [\href{https://competitions.codalab.org/competitions/17094}{link}] & 1 & 131 & 7 \\
    14. AbdomenCT-1K \citeyearpar{ma2021abdomenct} [\href{https://github.com/JunMa11/AbdomenCT-1K}{link}] & 4 & 1,050 & 12 & 15. KiTS \citeyearpar{heller2020international} [\href{https://kits-challenge.org/kits23/}{link}] & 1 & 489 & 1 \\
    16. FLARE'23 \citeyearpar{ma2022fast} [\href{https://codalab.lisn.upsaclay.fr/competitions/12239}{link}] & 13 & 4,100 & 30 & 17. Trauma Det.~\citeyearpar{rsna-2023-abdominal-trauma-detection} [\href{https://www.rsna.org/education/ai-resources-and-training/ai-image-challenge/abdominal-trauma-detection-ai-challenge}{link}] & 0 & 4,711 & 23 \\
    \hline
    18. AbdomenAtlas 1.0~\citeyearpar{qu2023annotating} [\href{https://github.com/MrGiovanni/AbdomenAtlas}{link}] & 9 & 5,195 & 26 & 19. \ourdataset & 25 & \totalct$^{\ddagger}$ & 88  \\
    \bottomrule
    \end{tabular}
    \begin{tablenotes}
        \item $^{\dagger}$Our reported number of CT volumes may differ from original publications, as some CT volumes are reserved for validation purposes.
        \item $^{\ddagger}$The number of CT volumes in \ourdataset\ is lower than the sum of datasets 1--17 due to overlaps within these public datasets.
    \end{tablenotes}
    \label{tab:dataset_overview}
\end{table*}

Interactive segmentation, an integration of AI algorithms and human expertise, was used to create \ourdataset\ in a semi-automatic procedure. We recruited a team of ten radiologists to perform manual annotations to ensure the annotation quality\footnote{Ensuring high-quality annotations is costly and time-consuming, yet it is critical for transfer learning and reducing ambiguity when training AI models for image segmentation.}. Given the complexity of 3D data, rather than annotating the entire dataset voxel by voxel, we asked the radiologists to focus on the most important CT volumes and regions therein. In doing so, an importance score for each volume was computed, derived from the uncertainty, consistency, and overlap \citep{qu2023annotating}. Six junior radiologists revised the annotations predicted by AI under the supervision of four senior radiologists, and in turn, AI improved its predictions by learning from these revised annotations. This interactive procedure continued to enhance the quality of annotations until no major revision was required from the radiologists. Subsequently, four senior radiologists went through the final visualizations for all the annotations, detecting and revising major errors as needed before the dataset was released. Annotation tools employed included a licensed version from \href{https://aipair.com.cn/}{Pair} and an open-source \href{https://monai.io/label.html}{MONAI Label}.

\ourdataset\ is a composite dataset that unifies CT volumes from public datasets 1--17 as summarized in~\tableautorefname~\ref{tab:dataset_overview}. \ourdataset\ presents a level of diversity because the CT volumes are sourced from 88 hospitals worldwide, including pre, portal, arterial, and delayed phases. The gap between these CT volumes includes changes in image quality due to different acquisition parameters, reconstruction kernels, and contrast enhancement, shown in Appendix~\ref{sec:domain_gaps_appendix}.
Moreover, we provide per-voxel annotations for 25 anatomical structures, including 16 abdominal organs, two thorax organs, five vascular structures, and two skeletal structures. We also provide pseudo annotations for seven types of tumors, namely liver, kidneys, pancreatic, hepatic vessel, lung, colon tumors, and kidney cysts. In total, more than 272.7B voxels are annotated in \ourdataset, marking a significant leap compared with the 4.3B voxels annotated in the public datasets, amplifying the annotations by a factor of 63.4$\times$ (shown in Appendix~\figureautorefname~\ref{fig:dataset_completion}). The high annotation quality is due to the uniform annotation standards described in \appendixname\ \ref{sec:annotation_standard_appendix}.
\textit{We commit to releasing \ourdataset\ to the public.} However, this dataset, the largest public per-voxel annotated CT collection by far, accounts for around 0.01\% of the CT volumes annually acquired in the United States \citep{papanicolas2018health}. Therefore, cross-institutional collaboration is crucial for accelerating data sharing, annotation, and AI development \citep{saenz2024maida}.

\subsection{A Suite of Pre-trained Models: \ourmodel}\label{sec:model_zoo}

The magnitude of our \ourdataset\ is unprecedented in terms of data and annotations. One of the advantages is that it enables us to train AI models in both a supervised and self-supervised manner. At the time this paper is written, neither supervised nor self-supervised pre-training has been performed on this scale of dataset (\totalct\ volumetric data)\footnote{For supervised pre-training, the largest study to date was by \citet{liu2023clip}, which was developed on 3,410 (2,100 for training and 1,310 for validation) annotated CT volumes. For self-supervised pre-training, the largest one was by \citet{tang2022self}, which was trained on 5,050 unannotated CT volumes. Concurrently, \citet{valanarasu2023disruptive} pre-trained a model on 50K volumes of CT and MRI using self-supervised learning.}. We have developed models (termed \ourmodel) pre-trained on data and annotations in \ourdataset, which leverage established CNN backbones, such as U-Net and SegResNet, as well as Transformer backbones, such as Swin UNETR. With the growing trend of using pre-trained models, we have maintained a standardized, accessible \href{https://github.com/MrGiovanni/SuPreM#a-suite-of-pre-trained-models-suprem}{model repository} for sharing public model weights as well as a suite of supervised pre-trained models (\ourmodel) released by us. Releasing pre-trained models should be considered a marked contribution as they offer an alternative way of knowledge sharing while protecting patient privacy \citep{sellergren2022simplified,zhang2023challenges,ma2023foundation}. 
In this study, all of the models in \ourmodel\ follow pre-training and fine-tuning configurations as below.

\begin{figure}[h]
  \begin{minipage}[c]{0.43\textwidth}
    \includegraphics[width=\textwidth]{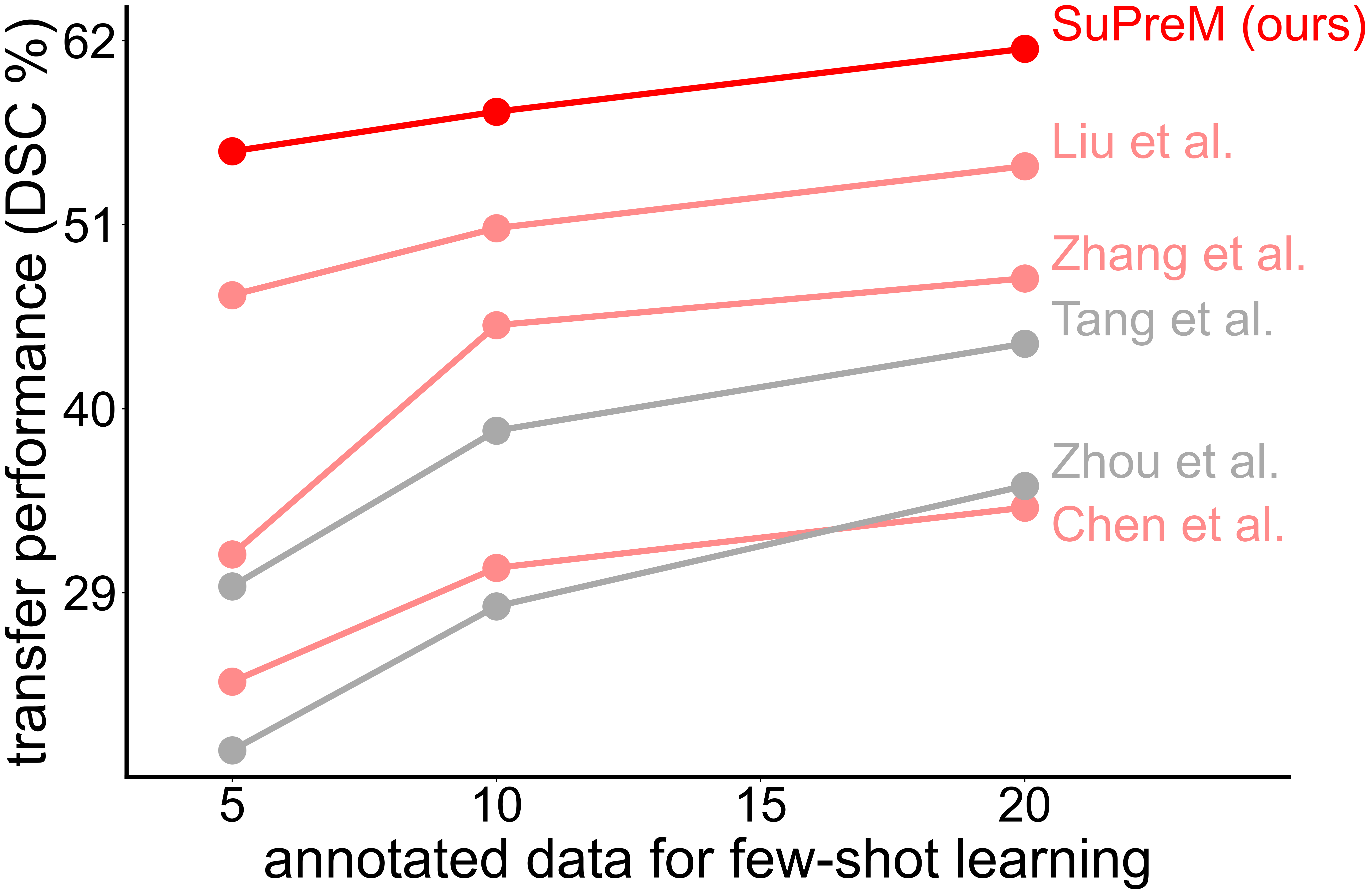}
  \end{minipage}\hfill
  \begin{minipage}[c]{0.55\textwidth}
    \caption{
    We present the transfer performance on a proprietary dataset with few-shot examples  ($N$ = $5,10,20$). The transfer performance (Y-axis) stands for the average DSC score across 20-class organ segmentation and 3-class tumor segmentation. Generally speaking, in a few-shot learning setting, supervised pre-trained models (in \textcolor{red}{red}) transfer better than self-supervised pre-trained models (in \textcolor{gray}{gray}). Notably, our \ourmodel\ achieves the best transfer performance over other well-known publicly available models.
    }
    \label{fig:few_shot}
  \end{minipage}
\end{figure}

To perform a fair and rigorous comparison, we benchmarked with public pre-training methods by pre-training \ourmodel\ using 2,100 CT volumes (same as \citet{liu2023clip} and fewer than \citet{tang2022self}) in Tables~\ref{tab:model_zoo}, \ref{tab:novel_classes} and Figures~\ref{fig:few_shot}, \ref{fig:data_compute_effciency}b, \ref{fig:pancreas_tumor_classification_task}. Then, we scaled up the number of CT volumes for pre-training to \totalct\ CT volumes to perform direct inference in \tableautorefname~\ref{tab:external_datasets}. Lastly, we scaled down the number of CT volumes to 21 to explore the edge of our \ourmodel\ in \figureautorefname~\ref{fig:data_compute_effciency}a. All these pre-trained models and configurations have been summarized in \appendixname\ \tableautorefname~\ref{tab:model_summary_appendix}. The best-performing model was selected based on the highest average DSC score over 32 classes on a validation set of 1,310 CT volumes. Implementation details of both pre-training and fine-tuning can be found in \appendixname\ \ref{sec:benchmark_results_appendix}.

\begin{table}[t]
    \centering
    \scriptsize
    \caption{\textbf{Contribution \#2: A suite of pre-trained models (termed \ourmodel) comprising several widely recognized AI models.} 
    We provide pre-trained AI models based on CNN, Transformer, and their hybrid versions, and more AI models will be added. Each model was supervised pre-trained on large datasets and per-voxel annotations from \ourdataset. Compared with learning from scratch and publicly available models, fine-tuning the models in \ourmodel\ consistently achieves state-of-the-art organ and tumor segmentation performance on two datasets. All of the results, including the mean and standard deviation (mean$\pm$s.d.) across ten trials. In addition, we have further performed an independent two-sample $t$-test between learning from scratch and fine-tuning models in our \ourmodel. The performance gain is statistically significant at the $P=0.05$ level, with highlighting in \textcolor{red!75}{light red}.}
    \begin{tabular}{p{0.14\linewidth}p{0.133\linewidth}P{0.08\linewidth}P{0.08\linewidth}P{0.08\linewidth}|P{0.08\linewidth}P{0.08\linewidth}P{0.08\linewidth}}
    \toprule
     & & \multicolumn{3}{c|}{TotalSegmentator v1} & \multicolumn{3}{c}{proprietary dataset} \\
    model \textcolor{gray}{(\# of param)} & pre-training & organ & muscle & cardiac & organ & gastro & cardiac \\
    \midrule
    \multirow{6}{*}{\makecell[l]{U-Net~\citeyearpar{ronneberger2015u}\\family\\ \textcolor{gray}{(19.08M)}}} & scratch & 88.9{\tiny$\pm$0.6} & 92.9{\tiny$\pm$0.4} & 88.8{\tiny$\pm$0.7} & 85.6{\tiny$\pm$0.5
    } & 69.8{\tiny$\pm$1.2} & 38.1{\tiny$\pm$1.1}\\
     & \citet{zhou2019models} & 87.8 & 90.1 & 86.3 & 80.1 & 65.5 & 36.9 \\
     & \citet{chen2019med3d} & 86.9 & 91.4 & 87.4 & 79.0 & 66.2 & 36.7 \\
     & \citet{xie2022unimiss} & 88.5 & 92.9 & 89.0 & - & - & -\\
     & \citet{zhang2021dodnet} & 89.3 & 93.8 & 89.1 & 85.7 & 72.7 & 38.3\\
     & \textbf{\ourmodel} & \cellcolor{red!25}92.1{\tiny$\pm$0.3} & \cellcolor{red!25}95.4{\tiny$\pm$0.1} & \cellcolor{red!25}92.2{\tiny$\pm$0.3} & \cellcolor{red!25}90.8{\tiny$\pm$0.2} & \cellcolor{red!25}76.2{\tiny$\pm$0.8} & \cellcolor{red!25}70.5{\tiny$\pm$0.5}\\ 
    \midrule
    \multirow{4}{*}{\makecell[l]{Swin UNETR~\citeyearpar{hatamizadeh2021swin}\\ \textcolor{gray}{(62.19M)}}} & scratch & 86.4{\tiny$\pm$0.5} & 88.8{\tiny$\pm$0.5}& 84.5{\tiny$\pm$0.6}& 77.3{\tiny$\pm$0.9}& 65.9{\tiny$\pm$ 1.7}& 35.5{\tiny$\pm$1.4}\\
     & \citet{tang2022self} & 89.3 & 93.8 & 88.3 & 87.9 & 72.5 & 38.9\\
     & \citet{liu2023clip} & 89.7 & 94.1 & 89.4 & 89.1 & 74.6 & 67.6\\
     & \textbf{\ourmodel} & \cellcolor{red!25}91.3{\tiny$\pm$0.3}& \cellcolor{red!25}94.6{\tiny$\pm$0.2}& \cellcolor{red!25}90.3{\tiny$\pm$0.3}& \cellcolor{red!25}90.4{\tiny$\pm$0.7} & \cellcolor{red!25}75.9{\tiny$\pm$1.2}& \cellcolor{red!25}69.8{\tiny$\pm$0.9}\\ 
    \midrule
    \multirow{2}{*}{\makecell[l]{SegResNet~\citeyearpar{myronenko20193d}\\ \textcolor{gray}{(4.7M)}}} & scratch & 88.6{\tiny$\pm$0.5}& 91.3{\tiny$\pm$0.4} & 89.8{\tiny$\pm$0.4} & 80.6{\tiny$\pm$0.8} & 67.0{\tiny$\pm$1.4} & 36.0{\tiny$\pm$1.3}\\
     & \textbf{\ourmodel} & \cellcolor{red!25}91.3{\tiny$\pm$0.5}& \cellcolor{red!25}94.0{\tiny$\pm$0.1}& \cellcolor{red!25}91.3{\tiny$\pm$0.5} & \cellcolor{red!25}86.6{\tiny$\pm$0.3} & \cellcolor{red!25}73.7{\tiny$\pm$1.0} & \cellcolor{red!25}67.9{\tiny$\pm$0.8}\\ 
    \bottomrule
    \end{tabular}
    \label{tab:model_zoo}
\end{table}

The transfer learning ability is assessed by segmentation performance on two datasets, i.e., TotalSegmentator v1 and a proprietary dataset. Benchmarking results in \tableautorefname~\ref{tab:model_zoo} indicate that, in comparison with learning from scratch and with existing public models, those fine-tuned from our \ourmodel\ consistently attain superior organ, muscle, cardiac, and gastro segmentation performance on both datasets. U-Net, as a simple and lightweight segmentation backbone, still performs competitively compared with alternative choices like Swin UNETR. This observation is aligned with the majority of the medical imaging community \citep{isensee2021nnu,eisenmann2023winner}, suggesting that more exploration is needed for advancing segmentation backbones. Moreover, in the scenarios of either small data regimes shown in \figureautorefname~\ref{fig:few_shot} or large data regimes shown in \appendixname\ \figureautorefname~\ref{fig:comprehensive_benchmark_appendix}a--d, supervised models transfer better than their self-supervised counterparts. In summary, our \ourmodel\ surpasses all existing 3D pre-trained models by a large margin in transfer learning performance, irrespective of their pre-training methodologies or data sources.

\section{Experiment \& Analysis}\label{sec:experiment_analysis}

\subsection{Data, Annotation, and Computational Efficiency}\label{sec:data_compute_efficiency}

\textbf{\textit{Summary.}} We demonstrate the remarkable efficiency: (1) \ourmodel\ trained with 21 CT volumes, 672 masks, and 40 GPU hours shows transfer learning ability similar to that trained with 5,050 CT volumes and 1,152 GPU hours. (2) \ourmodel\ requires 50\% fewer manual annotations for organ/tumor segmentation than self-supervised pre-training.

\begin{figure}[t]
\centering
\includegraphics[width=1.0\columnwidth]{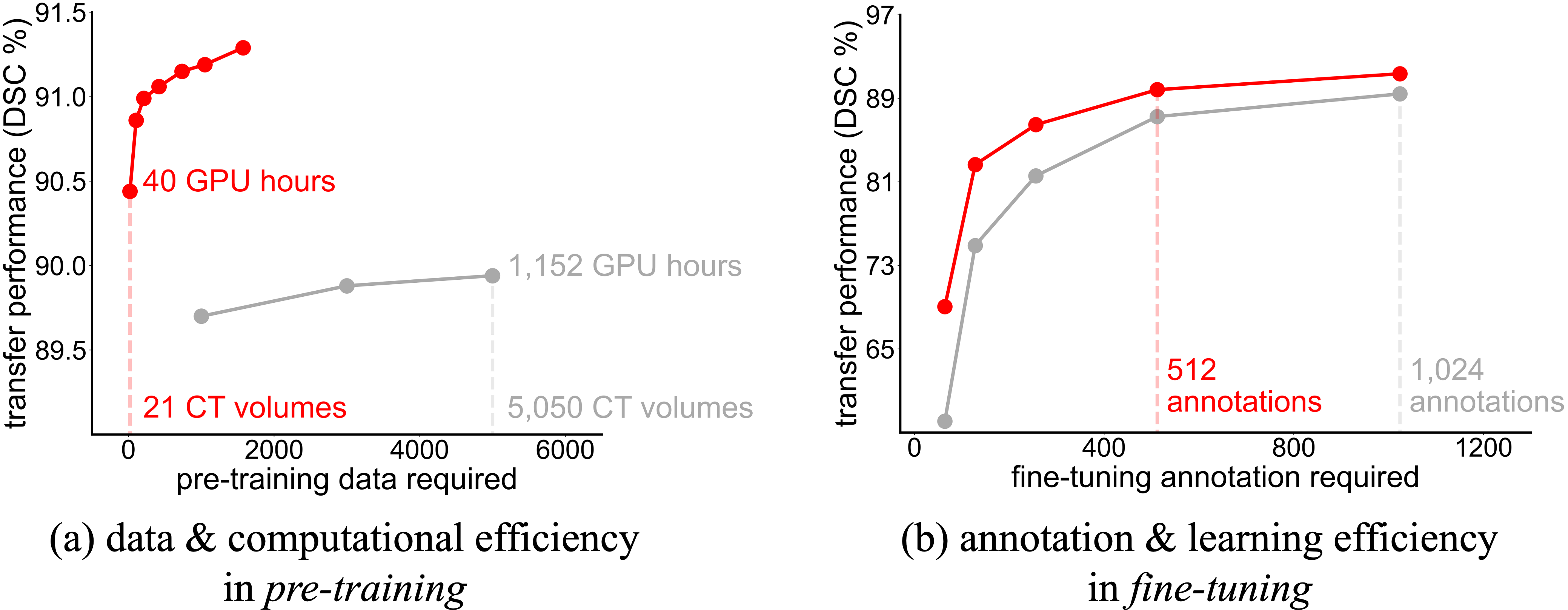}
    \caption{
    \textbf{Analysis of pre-training and fine-tuning efficiency.}
    For a fair comparison, both supervised (in \textcolor{red}{red}) and self-supervised (in \textcolor{gray}{gray}) models use Swin UNETR as the backbone, and the compared self-supervised pre-training is the current state of the art \citep{tang2022self}. The target task was on TotalSegmentator v1.
    \textbf{(a)} scales the model transfer learning ability when pre-trained on varying numbers of images. The results indicate a consistent improvement in transfer learning ability when pre-training on more images. The model trained with 21 CT volumes, 672 masks, and 40 GPU hours shows a transfer learning ability similar to that trained with 5,050 CT volumes and 1,152 GPU hours. Specifically, supervised pre-training is more efficient, requiring 99.6\% fewer data and 96.5\% less computation. 
    \textbf{(b)} assesses the annotation \& learning efficiency by fine-tuning models on different numbers of annotated CT volumes from TotalSegmentator. Specifically, \ourmodel, fine-tuned on 512 per-voxel annotated CT volumes, can achieve a segmentation performance on par with self-supervised models fine-tuned on 1,024 volumes, reducing 50\% manual annotation cost for target tasks.} 
\label{fig:data_compute_effciency}
\end{figure}

\textbf{\textit{Data efficiency}} for pre-training. As shown in~\figureautorefname~\ref{fig:data_compute_effciency}a, supervised pre-training requires less data (21 vs. 5,050 CT volumes) for the pretext task than self-supervised pre-training. This discrepancy arises from the inherent differences in their learning learning objectives and the information they leverage. Supervised pre-training benefits from explicit annotations, which provide direct guidance for the task, i.e., segmentation in this study. The model learns features from both data and annotations, which offer strong and precise supervision. On the other hand, self-supervised learning relies on pretext tasks derived from the raw data, which may offer a more ambiguous learning signal, therefore requiring more examples to capture meaningful features. Importantly, our finding suggests that supervised pre-training is more scalable with increased data. When data are increased from 21 to 1,575 volumes, the transfer learning performance on TotalSegmentator improves from 90.4\% to 91.3\%. In comparison, for self-supervised pre-training, an increase in data from 1,000 to 5,050 volumes only marginally improves performance from 89.7\% to 89.9\%. Therefore, supervised pre-training requires significantly less data than self-supervised and is more scalable and effective with increased data.

\textbf{\textit{Annotation efficiency}} for fine-tuning. We have assessed the annotation efficiency by fine-tuning \ourmodel\ and self-supervised models \citep{tang2022self} on the TotalSegmentator dataset. \figureautorefname~\ref{fig:data_compute_effciency}b suggests that fine-tuning \ourmodel\ can reduce annotation costs for the segmentation task by 50\%, averaged over the classes that were not used for pre-training (per-class performance can be found in \appendixname\ \figureautorefname~\ref{fig:all_annotation_efficiency}a--d). Specifically, \ourmodel\ fine-tuned on 512 per-voxel annotated CT volumes can achieve segmentation performance similar to \citet{tang2022self} fine-tuned on 1,024 annotated CT volumes. The fine-tuning performance improvement gets bigger when the number of annotated CT volumes is limited in the target task (e.g., 64, 128, 256). In addition, similar levels of annotation efficiency (reduced 50\% cost) are observed when fine-tuning \ourmodel\ on the three-class tumor segmentation task using the proprietary dataset, as presented in \appendixname\ \figureautorefname~\ref{fig:all_annotation_efficiency}e--g.

\textbf{\textit{Computational efficiency}} for both pre-training and fine-tuning. This efficiency stems, in part, from the reduced data requirements inherent to supervised pre-training, as discussed above. As shown in \figureautorefname~\ref{fig:data_compute_effciency}a, supervised pre-training only needs 40 GPU hours to achieve a transfer learning performance comparable to that of self-supervised pre-training, which requires 1,152 GPU hours---a factor increase of 28.8$\times$. When fine-tuning on target tasks, such as on a 10\% subset of TotalSegmentator in \appendixname\ \figureautorefname~\ref{fig:computational_efficiency_appendix}, the supervised pre-trained model converges much faster than the self-supervised one, reducing the GPU hours needed from 60 to 20. This implies that image features learned by supervised pre-training are intrinsically more expressive, enabling the model to seamlessly adapt across a myriad of 3D image segmentation tasks with minimal annotated data for fine-tuning. This computational efficiency makes supervised pre-training a compelling choice for 3D image segmentation without compromising model performance, especially when the large, annotated dataset is available.

\subsection{Enhanced Features for Novel Datasets, Classes, and Tasks}\label{sec:semantic_features}

\textbf{\textit{Summary.}} The learned features manifest considerable generalizability and adaptability. The features can \textit{direct inference} for organ segmentation on external datasets of CT volumes taken from different hospitals. The features can also be \textit{fine-tuned} to segment novel organ/tumor classes and classify tumor sub-types with higher accuracy and less annotated data than those learned by self-supervision.

\begin{table}[t]
    \centering
    \scriptsize
    \caption{
    \textbf{Direct inference on three external datasets.} 
    We conduct external validation across four hospitals worldwide. Specifically, our \ourmodel---trained on \totalct\ CT volumes---is directly inferred on three external datasets, i.e., TotalSegmentator (representing the Central European population from Switzerland; one hospital), DAP~Atlas (the Central European population from Germany; two hospitals), and the proprietary dataset (the North American population from the United States; one hospital) measured by DSC scores. For every dataset, we compare the \textit{out-of-distribution} (OOD) performance obtained by \ourmodel\ with \textit{independently and identically distributed} (IID) performance obtained by AI models directly trained on that specific dataset, which are often considered as upper bound performance in domain transfer literature. We find that \ourmodel\ can be generalized well across external datasets without additional fine-tuning, yielding comparable or even superior performance to the IID counterparts, evidenced by the one-sample $t$-test results. \appendixname\ \ref{sec:direct_inference_appendix} provides visual examples of anatomical structure segmentation.
    }
    \begin{tabular}{p{0.085\linewidth}R{0.075\linewidth}@{ }L{0.03\linewidth}P{0.11\linewidth}|R{0.075\linewidth}@{ }L{0.03\linewidth}P{0.11\linewidth}|R{0.075\linewidth}@{ }L{0.03\linewidth}P{0.12\linewidth}}
    \toprule
    \multirow{2}{*}{class} & \multicolumn{3}{c|}{TotalSegmentator v1} & \multicolumn{3}{c|}{DAP~Atlas} & \multicolumn{3}{c}{our proprietary dataset} \\
     & \multicolumn{2}{c}{\ourmodel\ (OOD)} & \citeauthor{liu2023clip} (IID) & \multicolumn{2}{c}{\ourmodel\ (OOD)} & \citeauthor{jaus2023towards} (IID) & \multicolumn{2}{c}{\ourmodel\ (OOD))} & \citeauthor{wang2019abdominal} (IID) \\
    \midrule
    spleen & 96.0{\tiny$\pm$0.0} & $^{****}$ & 93.6 & 96.8{\tiny$\pm$0.0} & $^{\mathrm{ns}}$ & 96.8 & 95.0{\tiny$\pm$0.0} & $^{****}$ & 89.6 \\
    kidney right & 93.3{\tiny$\pm$0.1} & $^{*}$ & 94.1 & 96.3{\tiny$\pm$0.1} & $^{****}$ & 95.3 & 92.2{\tiny$\pm$0.0} & $^{****}$ & 88.0\\
    kidney left & 91.2{\tiny$\pm$0.2} & $^{****}$ & 87.7 & 96.4{\tiny$\pm$0.1} & $^{****}$ & 97.4 & 91.6{\tiny$\pm$0.1} & $^{****}$ & 83.9 \\
    gall bladder & 81.8{\tiny$\pm$0.3} & $^{****}$ & 73.9 & 87.6{\tiny$\pm$0.4} & $^{****}$ & 71.2 & 83.6{\tiny$\pm$0.2} & $^{\mathrm{ns}}$ & 85.4\\
    liver & 96.4{\tiny$\pm$0.1} & $^{\mathrm{ns}}$ & 96.8 & 97.3{\tiny$\pm$0.1} & $^{****}$ & 98.5 & 95.0{\tiny$\pm$0.3} & $^{****}$ & 91.4\\
    stomach & 87.3{\tiny$\pm$0.3} & $^{\mathrm{ns}}$ & 89.2 & 95.3{\tiny$\pm$0.2} & $^{****}$ & 96.1 & 92.2{\tiny$\pm$0.1} & $^{*}$ & 93.6\\
    aorta & 80.8{\tiny$\pm$0.4} & $^{****}$ & 90.7 & 90.7{\tiny$\pm$0.5} & $^{****}$ & 97.7 & 73.9{\tiny$\pm$0.3} & $^{****}$ & 87.0\\
    postcava & 77.9{\tiny$\pm$0.3} & $^{****}$ & 82.1 & 89.1{\tiny$\pm$0.4} & $^{****}$ & 95.9 & 77.7{\tiny$\pm$0.4} & $^{**}$ & 80.8\\
    pancreas & 84.6{\tiny$\pm$0.2} & $^{****}$ & 80.8 & 90.6{\tiny$\pm$0.2} & $^{****}$ & 93.7 & 79.0{\tiny$\pm$0.3} & $^{\mathrm{ns}}$ & 79.3\\
    \hline
    \textbf{average} & 87.7{\tiny$\pm$0.2} & $^{\mathrm{ns}}$ & 87.6 & 93.3{\tiny$\pm$0.2} & $^{****}$ & 93.6 & 86.7{\tiny$\pm$0.2} & $^{\mathrm{ns}}$ & 86.1 \\
    \bottomrule
    \end{tabular}
    \begin{tablenotes}
        \item $^{\mathrm{ns}}$ $P>0.05$ \quad\quad $^{*}$ $P\leq0.05$ \quad\quad $^{**}$ $P\leq0.01$ \quad\quad $^{***}$ $P\leq0.001$ \quad\quad $^{****}$ $P\leq0.0001$
    \end{tablenotes}
    \label{tab:external_datasets}
\end{table}

\textbf{\textit{Direct inference on external datasets.}}
AI models trained on a specific dataset often encounter challenges in generalizing to novel datasets when a marked difference---referred to as a \textit{domain gap}---exists between them \citep{zhang2023challenges}. While domain adaptation and generalization are prevalent research strategies to mitigate this challenge \citep{guan2021domain,zhou2022domain}, we choose to address this issue by training a model on an expansive and diverse dataset (elaborated in \appendixname\ \ref{sec:domain_gaps_appendix}). We assume the domain gap between CT volumes from different hospitals is not as pronounced as those in computer vision. This is because of the relatively standardized nature of computer tomography as an imaging modality, where pixel intensity conveys consistent anatomical significance \citep{zhou2022interpreting}. \ourdataset\ presents impressive diversity, covering CT volumes with variations in contrast enhancement, reconstruction kernels, CT scanner types, and acquisition parameters. This breadth and diversity are imperative for developing an AI model with the robustness required to accommodate the variations present in novel datasets. 
We conduct external validation on several novel datasets sourced from Switzerland and East Asia to challenge the AI model on the data distribution that it has not encountered during the training. This result is referred to as \textit{out-of-distribution} (OOD) performance. For comparison, we also collect the result achieved by dataset-specific AI models---those individually trained on the specific datasets---referred to as \textit{independently and identically distributed} (IID) performance. As shown in \tableautorefname~\ref{tab:external_datasets}, our \ourmodel\ can be generalized well to novel data distribution without the need for further fine-tuning or adaptation, consistently offering OOD performance that matches or even exceeds that of its IID counterparts.

\begin{table}[t]
\caption{\textbf{Fine-tuning \ourmodel\ on 66 novel classes.}
    Following the standard transfer learning paradigm, we fine-tune our \ourmodel\ on the segmentation task of novel classes. These tasks include segmenting 19 muscles, 15 cardiac structures, 5 organs, and 24 vertebrae from TotalSegmentator, as well as three fine-grained pancreatic tumor types from the proprietary dataset. 
    It is important to note that these classes were not part of the pre-training of \ourmodel. We observe that \ourmodel, supervised pre-trained on only a few classes, can transfer better than those self-supervised pre-trained on raw, unlabeled data measured by DSC scores. In other words, it is the task of segmentation itself that can enhance the model's capability of segmenting novel-class objects. This benefit is much more straightforward and understandable than such self-supervised tasks as contextual prediction, mask image modeling, and instance discrimination in the context of transfer learning. We hypothesize that it is because the model learns to understand the concept of \textit{objectness} in a broader sense through full supervision, as suggested by \citet{kirillov2023segment}, but this certainly deserves further exploration. In addition, an independent two-sample $t$-test was performed between the self-supervised pre-trained model and the supervised pre-trained model.
    }
    \centering
    \scriptsize
    \begin{tabular}{p{0.15\linewidth}P{0.08\linewidth}R{0.08\linewidth}@{ }L{0.03\linewidth}P{0.03\linewidth}|p{0.155\linewidth}P{0.08\linewidth}R{0.08\linewidth}@{ }L{0.03\linewidth}P{0.03\linewidth}}
    \toprule
    novel class & self-super. & \multicolumn{2}{c}{super.} & $\Delta$ & novel class & self-super. & \multicolumn{2}{c}{super.} & $\Delta$ \\ 
    \midrule
    humerus left & 92.8{\tiny$\pm$0.7} & 93.2{\tiny$\pm$0.3} & $^{\mathrm{ns}}$ & 0.4 & vertebrae L5 & 94.1{\tiny$\pm$0.2} & 95.7{\tiny$\pm$0.3} & $^{****}$ & 1.6\\
    humerus right & 87.5{\tiny$\pm$1.0} & 95.0{\tiny$\pm$0.5} & $^{****}$ & 7.6 & vertebrae L4 & 90.4{\tiny$\pm$0.6} & 93.0{\tiny$\pm$0.5} & $^{****}$ & 2.6\\
    \multicolumn{5}{l|}{\textcolor{gray}{$\cdots$ (15 more classes)}} & \multicolumn{5}{l}{\textcolor{gray}{$\cdots$ (20 more classes)}}  \\
    iliopsoas left & 84.4{\tiny$\pm$0.3} & 85.7{\tiny$\pm$0.3} & $^{****}$ & 1.3 & vertebrae C2 & 86.8{\tiny$\pm$2.0} & 91.8{\tiny$\pm$0.2} & $^{****}$ & 5.1 \\
    iliopsoas right & 87.4{\tiny$\pm$0.3} & 88.7{\tiny$\pm$0.2} & $^{****}$ & 1.3 & vertebrae C1 & 87.1{\tiny$\pm$0.8} & 87.4{\tiny$\pm$0.8} & $^{\mathrm{ns}}$ & 0.3	\\
    \hline
    \textbf{average (muscle)} & 93.9{\tiny$\pm$0.1} & 94.3{\tiny$\pm$0.1} & $^{****}$ & 0.4 & \textbf{average (vertebrae)} & 86.4{\tiny$\pm$0.3} & 89.2{\tiny$\pm$0.2} & $^{****}$ & 2.7 \\  
     &  &  &  &  &  &  &  \\
    trachea & 93.4{\tiny$\pm$0.1} & 93.4{\tiny$\pm$0.1} & $^{\mathrm{ns}}$ & 0.0 \\
    heart myocardium & 88.9{\tiny$\pm$0.2} & 89.8{\tiny$\pm$0.2} & $^{****}$ & 0.9 & 	\\
    \multicolumn{5}{l|}{\textcolor{gray}{$\cdots$ (11 more classes)}}  & PDAC & 53.3{\tiny$\pm$0.4} & 53.6{\tiny$\pm$0.3} & $^{*}$ & 0.3\\
    urinary bladder & 90.5{\tiny$\pm$0.9} & 91.5{\tiny$\pm$0.9} & $^{*}$ & 1.0  &  Cyst & 41.5{\tiny$\pm$0.3} & 49.4{\tiny$\pm$0.3} & $^{****}$ & 7.9\\
    pulmonary artery & 89.0{\tiny$\pm$0.9} & 92.0{\tiny$\pm$0.2} & $^{****}$ & 3.0 & PanNet & 35.5{\tiny$\pm$0.8} & 46.0{\tiny$\pm$0.5} & $^{****}$ & 10.5\\ 
    \hline
    \textbf{average (cardiac)} & 88.9{\tiny$\pm$0.1} & 90.7{\tiny$\pm$0.1} & $^{****}$ & 1.8 & \textbf{average (tumor)} & 43.4{\tiny$\pm$0.3} & 49.7{\tiny$\pm$0.2} & $^{****}$ & 6.2 \\ 
    \bottomrule
\end{tabular}
\begin{tablenotes}
    \item $^{\mathrm{ns}}$ $P>0.05$ \quad\quad $^{*}$ $P\leq0.05$ \quad\quad $^{**}$ $P\leq0.01$ \quad\quad $^{***}$ $P\leq0.001$ \quad\quad $^{****}$ $P\leq0.0001$
\end{tablenotes}
\label{tab:novel_classes}
\end{table}

\textbf{\textit{Fine-tuning on novel classes.}}
The value of transfer learning lies in fine-tuning the pre-trained models on novel scenarios \citep{zhou2021models}, such as novel classes, image modalities, and vision tasks that are completely unseen during the pre-training. In this study, we evaluate the proficiency of \ourmodel\ when transferred to a wide variety of novel classes for 3D image segmentation tasks\footnote{The fine-tuning performance of 17 seen classes is promising, but this is expected because the model is exposed to more examples of these classes in both pre-training and fine-tuning phases.}. These novel classes include 19 muscles, 15 cardiac structures, 5 organs, and 24 vertebrae from the TotalSegmentator dataset, as well as three fine-grained pancreatic tumor types from the proprietary dataset. As shown in \tableautorefname~\ref{tab:novel_classes}, our \ourmodel, supervised pre-trained on 25 classes, can transfer better to novel classes than those self-supervised models pre-trained on raw, unlabeled data.
We find that the pretext task of segmentation itself can enhance the model capability of segmenting novel classes. The benefit of same-task transfer learning, i.e., segmentation as pretext and target tasks, is much more straightforward and understandable than other pretext tasks such as contextual prediction, mask image modeling, and instance discrimination. Through full supervision in segmentation tasks, the model learns to understand the concept of \textit{objectness}\footnote{Objectness refers to the inherent attributes that distinguish something as an object within an image, differentiating it from the background or other entities.}, wherein the model gains a more profound understanding of what characterizes an object. The model does not just recognize predefined objects but begins to understand the foundational factors of objects in general. Such factors include texture, boundary, shape, size, and other low-level visual cues that are often deemed essential for image segmentation. This resonates with our assertion in the introduction: just as classification-based features from ImageNet transfer optimally for classification tasks \citep{huh2016makes,he2019rethinking,zoph2020rethinking,ridnik2021imagenet}, segmentation-based features are optimal for segmentation tasks. Our findings do not negate the value of self-supervised pre-training. With \totalct\ CT volumes, should self-supervised pre-training outperforms supervised pre-training in model transferability in the future, its value will be further highlighted by eliminating the need for manual annotations.

\begin{figure}[t]
\centering
    \includegraphics[width=1.0\columnwidth]{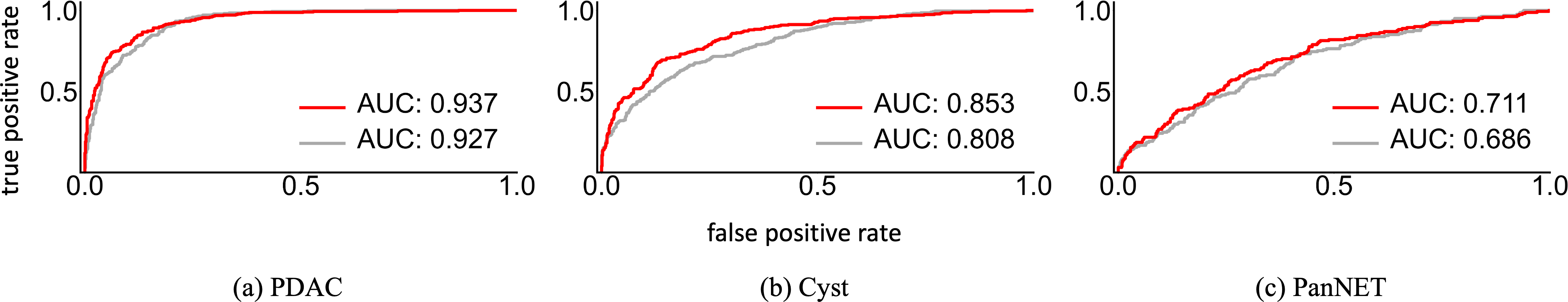}
    \caption{
    \textbf{Fine-tuning \ourmodel\ on fine-grained tumor classification.} We plot receiver operating characteristic (ROC) curves to evaluate the transfer learning performance of tumor classification. Detecting Cysts and PanNETs raises additional challenges for AI because these lesions exhibit a greater variety of texture patterns than PDACs. This diversity in texture patterns is reflected in the values of the Area Under the Curve (AUC) that we obtained. For all three sub-types of pancreatic tumors, \ourmodel\ (in \textcolor{red}{red}) demonstrates superior performance over the self-supervised model \citep{tang2022self} (in \textcolor{gray}{gray}), showcasing its effectiveness in fine-grained tumor classification.
    }
\label{fig:pancreas_tumor_classification_task}
\end{figure}

\textbf{\textit{Fine-tuning on novel tasks.}}
We have investigated the cross-task transfer learning ability of \ourmodel\ between organ segmentation and fine-grained tumor classification. The distance between the two tasks is much larger than transferring among segmentation tasks. It is challenging to benchmark fine-grained tumor classification, particularly due to the scarcity of annotations in public datasets (often limited to hundreds of tumors). To overcome this limitation, we employed our proprietary dataset \citep{xia2022felix}, which contains 1,869 and 1,073 CT volumes (i.e., 1,174 and 684 patients) for training and testing, respectively. These volumes present 3,577 annotated pancreatic tumors, including detailed sub-types: 1,704 PDACs, 945 Cysts, and 928 PanNETs. This extensive dataset enabled us to thoroughly assess the transfer learning ability of \ourmodel\ in tumor-related tasks.
\figureautorefname~\ref{fig:pancreas_tumor_classification_task} shows that supervised models (\ourmodel) transfer better to target classification tasks than self-supervised models \citep{tang2022self}, leading to improved Area Under the Curve (AUC) for identifying each tumor type.
Notably, the transfer learning results detailed in \appendixname\ \ref{sec:novel_task} reveal a sensitivity of 86.1\% and specificity of 95.4\% for PDAC detection. This performance surpasses the average radiologist's performance in PDAC identification by 27.6\% in sensitivity and 4.4\% in specificity, as reported in \citet{cao2023large}.
Moreover, \appendixname\ \figureautorefname~\ref{fig:all_annotation_efficiency} shows that \ourmodel\ requires 50\% fewer manual annotations for fine-grained tumor classification than self-supervised pre-training. This is particularly critical for tumor imaging tasks because annotating tumors requires much more effort and often relies on the availability of pathology reports.

\section{Conclusion and Discussion}\label{sec:conclusion}

This study examines the transfer learning ability of supervised models that are pre-trained on 3D annotated datasets and fine-tuned on 3D image segmentation tasks. We start by constructing \ourdataset, an extensive collection of \textbf{\totalct} three-dimensional CT volumes with high-quality, per-voxel annotations. The magnitude of this dataset is unprecedented regarding data volume (\textbf{\totalslices\ images}), granularity of annotations (\textbf{\totalmasks\ masks}), and inclusive diversity (\textbf{88 hospitals}). This dataset facilitates the development of a suite of pre-trained models, termed \ourmodel, that can be effectively transferred to a broad spectrum of 3D image segmentation tasks. Notably, \ourmodel\ transfers better than all existing 3D models by a large margin, especially when transferred to under-annotated datasets. The model trained with 21 CT volumes, 672 masks, and 40 GPU hours shows a transfer learning ability similar to that trained with 5,050 CT volumes and 1,152 GPU hours, highlighting the remarkable efficiency of supervised pre-training. We also demonstrate that the learned features can \textit{direct inference} effectively on external datasets and \textit{fine-tune} to segment novel classes and classify multiple types of tumors with higher accuracy and less annotated data than those learned by self-supervision.

\subsubsection*{Acknowledgments}
This work was supported by the Lustgarten Foundation for Pancreatic Cancer Research and the Patrick J. McGovern Foundation Award. This work has utilized the GPUs provided partially by ASU Research Computing and NVIDIA. We appreciate the effort of the MONAI Team to provide open-source code for the community. We thank Chongyu Qu, Yixiong Chen, Junfei Xiao, Jie Liu, Yucheng Tang, Tiezheng Zhang, Yaoyao Liu, Chen Wei, Fengrui Tian, Yu-Cheng Chou, Angtian Wang, and Dora Zhiyu Yang for their constructive suggestions at several stages of the project.

\bibliography{refs,zzhou}

\begin{thebibliography}{105}
\providecommand{\natexlab}[1]{#1}
\providecommand{\url}[1]{\texttt{#1}}
\expandafter\ifx\csname urlstyle\endcsname\relax
  \providecommand{\doi}[1]{doi: #1}\else
  \providecommand{\doi}{doi: \begingroup \urlstyle{rm}\Url}\fi

\bibitem[Antonelli et~al.(2021)Antonelli, Reinke, Bakas, Farahani, Landman,
  Litjens, Menze, Ronneberger, Summers, van Ginneken,
  et~al.]{antonelli2021medical}
Michela Antonelli, Annika Reinke, Spyridon Bakas, Keyvan Farahani, Bennett~A
  Landman, Geert Litjens, Bjoern Menze, Olaf Ronneberger, Ronald~M Summers,
  Bram van Ginneken, et~al.
\newblock The medical segmentation decathlon.
\newblock \emph{arXiv preprint arXiv:2106.05735}, 2021.

\bibitem[Bai et~al.(2023)Bai, Geng, Mangalam, Bar, Yuille, Darrell, Malik, and
  Efros]{bai2023sequential}
Yutong Bai, Xinyang Geng, Karttikeya Mangalam, Amir Bar, Alan Yuille, Trevor
  Darrell, Jitendra Malik, and Alexei~A Efros.
\newblock Sequential modeling enables scalable learning for large vision
  models.
\newblock \emph{arXiv preprint arXiv:2312.00785}, 2023.

\bibitem[Baxter et~al.(2023)Baxter, Nind, Sutherland, McAllister, Hardy, Hume,
  MacLeod, Caldwell, Krueger, Tramma, et~al.]{baxter2023scottish}
Rob Baxter, Thomas Nind, James Sutherland, Gordon McAllister, Douglas Hardy,
  Ally Hume, Ruairidh MacLeod, Jacqueline Caldwell, Susan Krueger, Leandro
  Tramma, et~al.
\newblock The scottish medical imaging archive: 57.3 million radiology studies
  linked to their medical records.
\newblock \emph{Radiology: Artificial Intelligence}, pp.\  e220266, 2023.

\bibitem[Bilic et~al.(2019)Bilic, Christ, Vorontsov, Chlebus, Chen, Dou, Fu,
  Han, Heng, Hesser, et~al.]{bilic2019liver}
Patrick Bilic, Patrick~Ferdinand Christ, Eugene Vorontsov, Grzegorz Chlebus,
  Hao Chen, Qi~Dou, Chi-Wing Fu, Xiao Han, Pheng-Ann Heng, J{\"u}rgen Hesser,
  et~al.
\newblock The liver tumor segmentation benchmark (lits).
\newblock \emph{arXiv preprint arXiv:1901.04056}, 2019.

\bibitem[Butoi et~al.(2023)Butoi, Ortiz, Ma, Sabuncu, Guttag, and
  Dalca]{butoi2023universeg}
Victor~Ion Butoi, Jose Javier~Gonzalez Ortiz, Tianyu Ma, Mert~R Sabuncu, John
  Guttag, and Adrian~V Dalca.
\newblock Universeg: Universal medical image segmentation.
\newblock \emph{arXiv preprint arXiv:2304.06131}, 2023.

\bibitem[Cao et~al.(2023)Cao, Xia, Yao, Han, Lambert, Zhang, Tang, Jin, Jiang,
  Fang, et~al.]{cao2023large}
Kai Cao, Yingda Xia, Jiawen Yao, Xu~Han, Lukas Lambert, Tingting Zhang, Wei
  Tang, Gang Jin, Hui Jiang, Xu~Fang, et~al.
\newblock Large-scale pancreatic cancer detection via non-contrast ct and deep
  learning.
\newblock \emph{Nature Medicine}, pp.\  1--11, 2023.

\bibitem[Chaitanya et~al.(2020)Chaitanya, Erdil, Karani, and
  Konukoglu]{chaitanya2020contrastive}
Krishna Chaitanya, Ertunc Erdil, Neerav Karani, and Ender Konukoglu.
\newblock Contrastive learning of global and local features for medical image
  segmentation with limited annotations.
\newblock \emph{arXiv preprint arXiv:2006.10511}, 2020.

\bibitem[Chen et~al.(2019{\natexlab{a}})Chen, Bentley, Mori, Misawa, Fujiwara,
  and Rueckert]{chen2019self}
Liang Chen, Paul Bentley, Kensaku Mori, Kazunari Misawa, Michitaka Fujiwara,
  and Daniel Rueckert.
\newblock Self-supervised learning for medical image analysis using image
  context restoration.
\newblock \emph{Medical image analysis}, 58:\penalty0 101539,
  2019{\natexlab{a}}.

\bibitem[Chen et~al.(2019{\natexlab{b}})Chen, Ma, and Zheng]{chen2019med3d}
Sihong Chen, Kai Ma, and Yefeng Zheng.
\newblock Med3d: Transfer learning for 3d medical image analysis.
\newblock \emph{arXiv preprint arXiv:1904.00625}, 2019{\natexlab{b}}.

\bibitem[Chen et~al.(2020{\natexlab{a}})Chen, Kornblith, Norouzi, and
  Hinton]{chen2020simple}
Ting Chen, Simon Kornblith, Mohammad Norouzi, and Geoffrey Hinton.
\newblock A simple framework for contrastive learning of visual
  representations.
\newblock \emph{arXiv preprint arXiv:2002.05709}, 2020{\natexlab{a}}.

\bibitem[Chen et~al.(2020{\natexlab{b}})Chen, Kornblith, Swersky, Norouzi, and
  Hinton]{chen2020big}
Ting Chen, Simon Kornblith, Kevin Swersky, Mohammad Norouzi, and Geoffrey~E
  Hinton.
\newblock Big self-supervised models are strong semi-supervised learners.
\newblock \emph{Advances in neural information processing systems},
  33:\penalty0 22243--22255, 2020{\natexlab{b}}.

\bibitem[Colak et~al.(2023)Colak, Lin, Ball, Davis, Flanders, Jalal, Magudia,
  Marinelli, Nicolaou, Prevedello, Rudie, Shih, Vazirabad, and
  Mongan]{rsna-2023-abdominal-trauma-detection}
Errol Colak, Hui-Ming Lin, Robyn Ball, Melissa Davis, Adam Flanders, Sabeena
  Jalal, Kirti Magudia, Brett Marinelli, Savvas Nicolaou, Luciano Prevedello,
  Jeff Rudie, George Shih, Maryam Vazirabad, and John Mongan.
\newblock Rsna 2023 abdominal trauma detection, 2023.
\newblock URL
  \url{https://kaggle.com/competitions/rsna-2023-abdominal-trauma-detection}.

\bibitem[Deng et~al.(2009)Deng, Dong, Socher, Li, Li, and
  Fei-Fei]{deng2009imagenet}
Jia Deng, Wei Dong, Richard Socher, Li-Jia Li, Kai Li, and Li~Fei-Fei.
\newblock Imagenet: A large-scale hierarchical image database.
\newblock In \emph{Proceedings of the IEEE Conference on Computer Vision and
  Pattern Recognition}, pp.\  248--255. IEEE, 2009.

\bibitem[Deng et~al.(2021)Deng, Wang, Hui, Li, Li, Luo, Sun, Quan, Yang, Hao,
  et~al.]{deng2021ctspine1k}
Yang Deng, Ce~Wang, Yuan Hui, Qian Li, Jun Li, Shiwei Luo, Mengke Sun, Quan
  Quan, Shuxin Yang, You Hao, et~al.
\newblock Ctspine1k: A large-scale dataset for spinal vertebrae segmentation in
  computed tomography.
\newblock \emph{arXiv preprint arXiv:2105.14711}, 2021.

\bibitem[Eisenmann et~al.(2023)Eisenmann, Reinke, Weru, Tizabi, Isensee, Adler,
  Ali, Andrearczyk, Aubreville, Baid, et~al.]{eisenmann2023winner}
Matthias Eisenmann, Annika Reinke, Vivienn Weru, Minu~D Tizabi, Fabian Isensee,
  Tim~J Adler, Sharib Ali, Vincent Andrearczyk, Marc Aubreville, Ujjwal Baid,
  et~al.
\newblock Why is the winner the best?
\newblock In \emph{Proceedings of the IEEE/CVF Conference on Computer Vision
  and Pattern Recognition}, pp.\  19955--19966, 2023.

\bibitem[Gatidis et~al.(2022)Gatidis, Hepp, Fr{\"u}h, La~Foug{\`e}re, Nikolaou,
  Pfannenberg, Sch{\"o}lkopf, K{\"u}stner, Cyran, and Rubin]{gatidis2022whole}
Sergios Gatidis, Tobias Hepp, Marcel Fr{\"u}h, Christian La~Foug{\`e}re,
  Konstantin Nikolaou, Christina Pfannenberg, Bernhard Sch{\"o}lkopf, Thomas
  K{\"u}stner, Clemens Cyran, and Daniel Rubin.
\newblock A whole-body fdg-pet/ct dataset with manually annotated tumor
  lesions.
\newblock \emph{Scientific Data}, 9\penalty0 (1):\penalty0 601, 2022.

\bibitem[Grill et~al.(2020)Grill, Strub, Altch{\'e}, Tallec, Richemond,
  Buchatskaya, Doersch, Avila~Pires, Guo, Gheshlaghi~Azar,
  et~al.]{grill2020bootstrap}
Jean-Bastien Grill, Florian Strub, Florent Altch{\'e}, Corentin Tallec, Pierre
  Richemond, Elena Buchatskaya, Carl Doersch, Bernardo Avila~Pires, Zhaohan
  Guo, Mohammad Gheshlaghi~Azar, et~al.
\newblock Bootstrap your own latent-a new approach to self-supervised learning.
\newblock \emph{Advances in neural information processing systems},
  33:\penalty0 21271--21284, 2020.

\bibitem[Guan \& Liu(2021)Guan and Liu]{guan2021domain}
Hao Guan and Mingxia Liu.
\newblock Domain adaptation for medical image analysis: a survey.
\newblock \emph{IEEE Transactions on Biomedical Engineering}, 69\penalty0
  (3):\penalty0 1173--1185, 2021.

\bibitem[Haghighi et~al.(2020)Haghighi, Hosseinzadeh~Taher, Zhou, Gotway, and
  Liang]{haghighi2020learning}
Fatemeh Haghighi, Mohammad~Reza Hosseinzadeh~Taher, Zongwei Zhou, Michael~B
  Gotway, and Jianming Liang.
\newblock Learning semantics-enriched representation via self-discovery,
  self-classification, and self-restoration.
\newblock In \emph{International Conference on Medical Image Computing and
  Computer-Assisted Intervention}, pp.\  137--147. Springer, 2020.
\newblock URL \url{https://github.com/fhaghighi/SemanticGenesis}.

\bibitem[Haghighi et~al.(2021)Haghighi, Taher, Zhou, Gotway, and
  Liang]{haghighi2021transferable}
Fatemeh Haghighi, Mohammad Reza~Hosseinzadeh Taher, Zongwei Zhou, Michael~B
  Gotway, and Jianming Liang.
\newblock Transferable visual words: Exploiting the semantics of anatomical
  patterns for self-supervised learning.
\newblock \emph{IEEE Transactions on Medical Imaging}, 2021.
\newblock URL \url{https://github.com/fhaghighi/SemanticGenesis}.

\bibitem[Hatamizadeh et~al.(2021)Hatamizadeh, Nath, Tang, Yang, Roth, and
  Xu]{hatamizadeh2021swin}
Ali Hatamizadeh, Vishwesh Nath, Yucheng Tang, Dong Yang, Holger~R Roth, and
  Daguang Xu.
\newblock Swin unetr: Swin transformers for semantic segmentation of brain
  tumors in mri images.
\newblock In \emph{International MICCAI Brainlesion Workshop}, pp.\  272--284.
  Springer, 2021.

\bibitem[He et~al.(2019)He, Girshick, and Doll{\'a}r]{he2019rethinking}
Kaiming He, Ross Girshick, and Piotr Doll{\'a}r.
\newblock Rethinking imagenet pre-training.
\newblock In \emph{Proceedings of the IEEE/CVF International Conference on
  Computer Vision}, pp.\  4918--4927, 2019.

\bibitem[He et~al.(2020)He, Fan, Wu, Xie, and Girshick]{he2020momentum}
Kaiming He, Haoqi Fan, Yuxin Wu, Saining Xie, and Ross Girshick.
\newblock Momentum contrast for unsupervised visual representation learning.
\newblock In \emph{Proceedings of the IEEE/CVF Conference on Computer Vision
  and Pattern Recognition}, pp.\  9729--9738, 2020.

\bibitem[He et~al.(2022)He, Chen, Xie, Li, Doll{\'a}r, and
  Girshick]{he2022masked}
Kaiming He, Xinlei Chen, Saining Xie, Yanghao Li, Piotr Doll{\'a}r, and Ross
  Girshick.
\newblock Masked autoencoders are scalable vision learners.
\newblock In \emph{Proceedings of the IEEE/CVF Conference on Computer Vision
  and Pattern Recognition}, pp.\  16000--16009, 2022.

\bibitem[He et~al.(2021)He, Yang, Yang, Ge, Kong, Zhu, Zhang, Shao, Shu,
  Dillenseger, et~al.]{he2021meta}
Yuting He, Guanyu Yang, Jian Yang, Rongjun Ge, Youyong Kong, Xiaomei Zhu,
  Shaobo Zhang, Pengfei Shao, Huazhong Shu, Jean-Louis Dillenseger, et~al.
\newblock Meta grayscale adaptive network for 3d integrated renal structures
  segmentation.
\newblock \emph{Medical image analysis}, 71:\penalty0 102055, 2021.

\bibitem[Heller et~al.(2020)Heller, McSweeney, Peterson, Peterson, Rickman,
  Stai, Tejpaul, Oestreich, Blake, Rosenberg, et~al.]{heller2020international}
Nicholas Heller, Sean McSweeney, Matthew~Thomas Peterson, Sarah Peterson, Jack
  Rickman, Bethany Stai, Resha Tejpaul, Makinna Oestreich, Paul Blake, Joel
  Rosenberg, et~al.
\newblock An international challenge to use artificial intelligence to define
  the state-of-the-art in kidney and kidney tumor segmentation in ct imaging.,
  2020.

\bibitem[Huang et~al.(2023)Huang, Wang, Deng, Ye, Su, Sun, He, Gu, Gu, Zhang,
  et~al.]{huang2023stu}
Ziyan Huang, Haoyu Wang, Zhongying Deng, Jin Ye, Yanzhou Su, Hui Sun, Junjun
  He, Yun Gu, Lixu Gu, Shaoting Zhang, et~al.
\newblock Stu-net: Scalable and transferable medical image segmentation models
  empowered by large-scale supervised pre-training.
\newblock \emph{arXiv preprint arXiv:2304.06716}, 2023.

\bibitem[Huh et~al.(2016)Huh, Agrawal, and Efros]{huh2016makes}
Minyoung Huh, Pulkit Agrawal, and Alexei~A Efros.
\newblock What makes imagenet good for transfer learning?
\newblock \emph{arXiv preprint arXiv:1608.08614}, 2016.

\bibitem[Ilharco et~al.(2022)Ilharco, Ribeiro, Wortsman, Gururangan, Schmidt,
  Hajishirzi, and Farhadi]{ilharco2022editing}
Gabriel Ilharco, Marco~Tulio Ribeiro, Mitchell Wortsman, Suchin Gururangan,
  Ludwig Schmidt, Hannaneh Hajishirzi, and Ali Farhadi.
\newblock Editing models with task arithmetic.
\newblock \emph{arXiv preprint arXiv:2212.04089}, 2022.

\bibitem[Isensee et~al.(2021)Isensee, Jaeger, Kohl, Petersen, and
  Maier-Hein]{isensee2021nnu}
Fabian Isensee, Paul~F Jaeger, Simon~AA Kohl, Jens Petersen, and Klaus~H
  Maier-Hein.
\newblock nnu-net: a self-configuring method for deep learning-based biomedical
  image segmentation.
\newblock \emph{Nature Methods}, 18\penalty0 (2):\penalty0 203--211, 2021.

\bibitem[Jaus et~al.(2023)Jaus, Seibold, Hermann, Walter, Giske, Haubold,
  Kleesiek, and Stiefelhagen]{jaus2023towards}
Alexander Jaus, Constantin Seibold, Kelsey Hermann, Alexandra Walter, Kristina
  Giske, Johannes Haubold, Jens Kleesiek, and Rainer Stiefelhagen.
\newblock Towards unifying anatomy segmentation: automated generation of a
  full-body ct dataset via knowledge aggregation and anatomical guidelines.
\newblock \emph{arXiv preprint arXiv:2307.13375}, 2023.

\bibitem[Ji et~al.(2022)Ji, Bai, Yang, Ge, Zhu, Zhang, Li, Zhang, Ma, Wan,
  et~al.]{ji2022amos}
Yuanfeng Ji, Haotian Bai, Jie Yang, Chongjian Ge, Ye~Zhu, Ruimao Zhang, Zhen
  Li, Lingyan Zhang, Wanling Ma, Xiang Wan, et~al.
\newblock Amos: A large-scale abdominal multi-organ benchmark for versatile
  medical image segmentation.
\newblock \emph{arXiv preprint arXiv:2206.08023}, 2022.

\bibitem[Jing \& Tian(2020)Jing and Tian]{jing2020self}
Longlong Jing and Yingli Tian.
\newblock Self-supervised visual feature learning with deep neural networks: A
  survey.
\newblock \emph{IEEE transactions on pattern analysis and machine
  intelligence}, 43\penalty0 (11):\penalty0 4037--4058, 2020.

\bibitem[Kirillov et~al.(2023)Kirillov, Mintun, Ravi, Mao, Rolland, Gustafson,
  Xiao, Whitehead, Berg, Lo, et~al.]{kirillov2023segment}
Alexander Kirillov, Eric Mintun, Nikhila Ravi, Hanzi Mao, Chloe Rolland, Laura
  Gustafson, Tete Xiao, Spencer Whitehead, Alexander~C Berg, Wan-Yen Lo, et~al.
\newblock Segment anything.
\newblock \emph{arXiv preprint arXiv:2304.02643}, 2023.

\bibitem[Kolesnikov et~al.(2020)Kolesnikov, Beyer, Zhai, Puigcerver, Yung,
  Gelly, and Houlsby]{kolesnikov2020big}
Alexander Kolesnikov, Lucas Beyer, Xiaohua Zhai, Joan Puigcerver, Jessica Yung,
  Sylvain Gelly, and Neil Houlsby.
\newblock Big transfer (bit): General visual representation learning.
\newblock In \emph{Computer Vision--ECCV 2020: 16th European Conference,
  Glasgow, UK, August 23--28, 2020, Proceedings, Part V 16}, pp.\  491--507.
  Springer, 2020.

\bibitem[Kumar(2017)]{kumar2017weight}
Siddharth~Krishna Kumar.
\newblock On weight initialization in deep neural networks.
\newblock \emph{arXiv preprint arXiv:1704.08863}, 2017.

\bibitem[Kuznetsova et~al.(2020)Kuznetsova, Rom, Alldrin, Uijlings, Krasin,
  Pont-Tuset, Kamali, Popov, Malloci, Kolesnikov, et~al.]{kuznetsova2020open}
Alina Kuznetsova, Hassan Rom, Neil Alldrin, Jasper Uijlings, Ivan Krasin, Jordi
  Pont-Tuset, Shahab Kamali, Stefan Popov, Matteo Malloci, Alexander
  Kolesnikov, et~al.
\newblock The open images dataset v4: Unified image classification, object
  detection, and visual relationship detection at scale.
\newblock \emph{International Journal of Computer Vision}, 128\penalty0
  (7):\penalty0 1956--1981, 2020.

\bibitem[Landman et~al.(2015)Landman, Xu, Igelsias, Styner, Langerak, and
  Klein]{landman2015miccai}
Bennett Landman, Zhoubing Xu, J~Igelsias, Martin Styner, T~Langerak, and Arno
  Klein.
\newblock Miccai multi-atlas labeling beyond the cranial vault--workshop and
  challenge.
\newblock In \emph{Proc. MICCAI Multi-Atlas Labeling Beyond Cranial
  Vault—Workshop Challenge}, volume~5, pp.\ ~12, 2015.

\bibitem[Li et~al.(2024)Li, Zhou, Yuille, Allan, and McLeod]{li2024ultra}
Bowen Li, Zongwei Zhou, Alan Yuille, Max Allan, and Jonathan McLeod.
\newblock Ultra-transunet: ultrasound segmentation framework with
  spatial-temporal context feature fusion.
\newblock In \emph{Medical Imaging 2024: Ultrasonic Imaging and Tomography},
  volume 12932, pp.\  8--15. SPIE, 2024.

\bibitem[Liu et~al.(2023)Liu, Zhang, Chen, Xiao, Lu, A~Landman, Yuan, Yuille,
  Tang, and Zhou]{liu2023clip}
Jie Liu, Yixiao Zhang, Jie-Neng Chen, Junfei Xiao, Yongyi Lu, Bennett
  A~Landman, Yixuan Yuan, Alan Yuille, Yucheng Tang, and Zongwei Zhou.
\newblock Clip-driven universal model for organ segmentation and tumor
  detection.
\newblock In \emph{Proceedings of the IEEE/CVF International Conference on
  Computer Vision}, pp.\  21152--21164, 2023.
\newblock URL \url{https://github.com/ljwztc/CLIP-Driven-Universal-Model}.

\bibitem[Luo et~al.(2021)Luo, Liao, Xiao, Song, Zhang, Li, Wang, and
  Zhang]{luo2021word}
Xiangde Luo, Wenjun Liao, Jianghong Xiao, Tao Song, Xiaofan Zhang, Kang Li,
  Guotai Wang, and Shaoting Zhang.
\newblock Word: Revisiting organs segmentation in the whole abdominal region.
\newblock \emph{arXiv preprint arXiv:2111.02403}, 2021.

\bibitem[Ma et~al.(2023{\natexlab{a}})Ma, Pang, Gotway, and
  Liang]{ma2023foundation}
DongAo Ma, Jiaxuan Pang, Michael~B Gotway, and Jianming Liang.
\newblock Foundation ark: Accruing and reusing knowledge for superior and
  robust performance.
\newblock In \emph{International Conference on Medical Image Computing and
  Computer-Assisted Intervention}, pp.\  651--662. Springer,
  2023{\natexlab{a}}.

\bibitem[Ma \& Wang(2023{\natexlab{a}})Ma and Wang]{ma2023segment}
Jun Ma and Bo~Wang.
\newblock Segment anything in medical images.
\newblock \emph{arXiv preprint arXiv:2304.12306}, 2023{\natexlab{a}}.

\bibitem[Ma \& Wang(2023{\natexlab{b}})Ma and Wang]{ma2023towards}
Jun Ma and Bo~Wang.
\newblock Towards foundation models of biological image segmentation.
\newblock \emph{Nature Methods}, 20\penalty0 (7):\penalty0 953--955,
  2023{\natexlab{b}}.

\bibitem[Ma et~al.(2021)Ma, Zhang, Gu, Zhu, Ge, Zhang, An, Wang, Wang, Liu,
  et~al.]{ma2021abdomenct}
Jun Ma, Yao Zhang, Song Gu, Cheng Zhu, Cheng Ge, Yichi Zhang, Xingle An,
  Congcong Wang, Qiyuan Wang, Xin Liu, et~al.
\newblock Abdomenct-1k: Is abdominal organ segmentation a solved problem.
\newblock \emph{IEEE Transactions on Pattern Analysis and Machine
  Intelligence}, 2021.

\bibitem[Ma et~al.(2022)Ma, Zhang, Gu, An, Wang, Ge, Wang, Zhang, Wang, Xu,
  et~al.]{ma2022fast}
Jun Ma, Yao Zhang, Song Gu, Xingle An, Zhihe Wang, Cheng Ge, Congcong Wang, Fan
  Zhang, Yu~Wang, Yinan Xu, et~al.
\newblock Fast and low-gpu-memory abdomen ct organ segmentation: the flare
  challenge.
\newblock \emph{Medical Image Analysis}, 82:\penalty0 102616, 2022.

\bibitem[Ma et~al.(2023{\natexlab{b}})Ma, Zhang, Gu, Ge, Ma, Young, Zhu, Meng,
  Yang, Huang, et~al.]{ma2023unleashing}
Jun Ma, Yao Zhang, Song Gu, Cheng Ge, Shihao Ma, Adamo Young, Cheng Zhu,
  Kangkang Meng, Xin Yang, Ziyan Huang, et~al.
\newblock Unleashing the strengths of unlabeled data in pan-cancer abdominal
  organ quantification: the flare22 challenge.
\newblock \emph{arXiv preprint arXiv:2308.05862}, 2023{\natexlab{b}}.

\bibitem[Ma et~al.(2023{\natexlab{c}})Ma, Li, Du, Zhang, Tang, Ma, Huang, Liu,
  Sun, Chen, et~al.]{ma2023aatct}
Zhiyu Ma, Chen Li, Tianming Du, Le~Zhang, Dechao Tang, Deguo Ma, Shanchuan
  Huang, Yan Liu, Yihao Sun, Zhihao Chen, et~al.
\newblock Aatct-ids: A benchmark abdominal adipose tissue ct image dataset for
  image denoising, semantic segmentation, and radiomics evaluation.
\newblock \emph{arXiv preprint arXiv:2308.08172}, 2023{\natexlab{c}}.

\bibitem[Mahajan et~al.(2018)Mahajan, Girshick, Ramanathan, He, Paluri, Li,
  Bharambe, and Van Der~Maaten]{mahajan2018exploring}
Dhruv Mahajan, Ross Girshick, Vignesh Ramanathan, Kaiming He, Manohar Paluri,
  Yixuan Li, Ashwin Bharambe, and Laurens Van Der~Maaten.
\newblock Exploring the limits of weakly supervised pretraining.
\newblock In \emph{Proceedings of the European conference on computer vision
  (ECCV)}, pp.\  181--196, 2018.

\bibitem[Masoudi et~al.(2018)Masoudi, Pourreza, Saadatmand-Tarzjan, Eftekhari,
  Zargar, and Rad]{masoudi2018new}
Mojtaba Masoudi, Hamid-Reza Pourreza, Mahdi Saadatmand-Tarzjan, Noushin
  Eftekhari, Fateme~Shafiee Zargar, and Masoud~Pezeshki Rad.
\newblock A new dataset of computed-tomography angiography images for
  computer-aided detection of pulmonary embolism.
\newblock \emph{Scientific data}, 5\penalty0 (1):\penalty0 1--9, 2018.

\bibitem[Mei et~al.(2022)Mei, Liu, Robson, Marinelli, Huang, Doshi, Jacobi,
  Cao, Link, Yang, et~al.]{mei2022radimagenet}
Xueyan Mei, Zelong Liu, Philip~M Robson, Brett Marinelli, Mingqian Huang, Amish
  Doshi, Adam Jacobi, Chendi Cao, Katherine~E Link, Thomas Yang, et~al.
\newblock Radimagenet: an open radiologic deep learning research dataset for
  effective transfer learning.
\newblock \emph{Radiology: Artificial Intelligence}, 4\penalty0 (5):\penalty0
  e210315, 2022.

\bibitem[Moor et~al.(2023)Moor, Banerjee, Abad, Krumholz, Leskovec, Topol, and
  Rajpurkar]{moor2023foundation}
Michael Moor, Oishi Banerjee, Zahra Shakeri~Hossein Abad, Harlan~M Krumholz,
  Jure Leskovec, Eric~J Topol, and Pranav Rajpurkar.
\newblock Foundation models for generalist medical artificial intelligence.
\newblock \emph{Nature}, 616\penalty0 (7956):\penalty0 259--265, 2023.

\bibitem[Myronenko(2019)]{myronenko20193d}
Andriy Myronenko.
\newblock 3d mri brain tumor segmentation using autoencoder regularization.
\newblock In \emph{Brainlesion: Glioma, Multiple Sclerosis, Stroke and
  Traumatic Brain Injuries: 4th International Workshop, BrainLes 2018, Held in
  Conjunction with MICCAI 2018, Granada, Spain, September 16, 2018, Revised
  Selected Papers, Part II 4}, pp.\  311--320. Springer, 2019.

\bibitem[Noroozi \& Favaro(2016)Noroozi and Favaro]{noroozi2016unsupervised}
Mehdi Noroozi and Paolo Favaro.
\newblock Unsupervised learning of visual representations by solving jigsaw
  puzzles.
\newblock In \emph{European Conference on Computer Vision}, pp.\  69--84.
  Springer, 2016.

\bibitem[Papanicolas et~al.(2018)Papanicolas, Woskie, and
  Jha]{papanicolas2018health}
Irene Papanicolas, Liana~R Woskie, and Ashish~K Jha.
\newblock Health care spending in the united states and other high-income
  countries.
\newblock \emph{Jama}, 319\penalty0 (10):\penalty0 1024--1039, 2018.

\bibitem[Park et~al.(2020)Park, Chu, Fishman, Yuille, Vogelstein, Kinzler,
  Horton, Hruban, Zinreich, Fouladi, et~al.]{park2020annotated}
S~Park, LC~Chu, EK~Fishman, AL~Yuille, B~Vogelstein, KW~Kinzler, KM~Horton,
  RH~Hruban, ES~Zinreich, D~Fadaei Fouladi, et~al.
\newblock Annotated normal ct data of the abdomen for deep learning: Challenges
  and strategies for implementation.
\newblock \emph{Diagnostic and interventional imaging}, 101\penalty0
  (1):\penalty0 35--44, 2020.

\bibitem[Pathak et~al.(2016)Pathak, Krahenbuhl, Donahue, Darrell, and
  Efros]{pathak2016context}
Deepak Pathak, Philipp Krahenbuhl, Jeff Donahue, Trevor Darrell, and Alexei~A
  Efros.
\newblock Context encoders: Feature learning by inpainting.
\newblock In \emph{Proceedings of the IEEE Conference on Computer Vision and
  Pattern Recognition}, pp.\  2536--2544, 2016.

\bibitem[Qu et~al.(2023)Qu, Zhang, Qiao, Liu, Tang, Yuille, and
  Zhou]{qu2023annotating}
Chongyu Qu, Tiezheng Zhang, Hualin Qiao, Jie Liu, Yucheng Tang, Alan Yuille,
  and Zongwei Zhou.
\newblock Abdomenatlas-8k: Annotating 8,000 abdominal ct volumes for
  multi-organ segmentation in three weeks.
\newblock In \emph{Conference on Neural Information Processing Systems},
  volume~21, 2023.
\newblock URL \url{https://github.com/MrGiovanni/AbdomenAtlas}.

\bibitem[Radford et~al.(2021)Radford, Kim, Hallacy, Ramesh, Goh, Agarwal,
  Sastry, Askell, Mishkin, Clark, et~al.]{radford2021learning}
Alec Radford, Jong~Wook Kim, Chris Hallacy, Aditya Ramesh, Gabriel Goh,
  Sandhini Agarwal, Girish Sastry, Amanda Askell, Pamela Mishkin, Jack Clark,
  et~al.
\newblock Learning transferable visual models from natural language
  supervision.
\newblock In \emph{International conference on machine learning}, pp.\
  8748--8763. PMLR, 2021.

\bibitem[Ren et~al.(2022)Ren, Wang, Gao, He, Yuille, Zhou, and
  Xie]{ren2022simple}
Sucheng Ren, Huiyu Wang, Zhengqi Gao, Shengfeng He, Alan Yuille, Yuyin Zhou,
  and Cihang Xie.
\newblock A simple data mixing prior for improving self-supervised learning.
\newblock In \emph{Proceedings of the IEEE/CVF conference on computer vision
  and pattern recognition}, pp.\  14595--14604, 2022.

\bibitem[Ren et~al.(2023)Ren, Wei, Zhang, and Hu]{ren2023tinymim}
Sucheng Ren, Fangyun Wei, Zheng Zhang, and Han Hu.
\newblock Tinymim: An empirical study of distilling mim pre-trained models.
\newblock In \emph{Proceedings of the IEEE/CVF Conference on Computer Vision
  and Pattern Recognition}, pp.\  3687--3697, 2023.

\bibitem[Ridnik et~al.(2021)Ridnik, Ben-Baruch, Noy, and
  Zelnik-Manor]{ridnik2021imagenet}
Tal Ridnik, Emanuel Ben-Baruch, Asaf Noy, and Lihi Zelnik-Manor.
\newblock Imagenet-21k pretraining for the masses.
\newblock \emph{arXiv preprint arXiv:2104.10972}, 2021.

\bibitem[Rister et~al.(2020)Rister, Yi, Shivakumar, Nobashi, and
  Rubin]{rister2020ct}
Blaine Rister, Darvin Yi, Kaushik Shivakumar, Tomomi Nobashi, and Daniel~L
  Rubin.
\newblock Ct-org, a new dataset for multiple organ segmentation in computed
  tomography.
\newblock \emph{Scientific Data}, 7\penalty0 (1):\penalty0 1--9, 2020.

\bibitem[Ronneberger et~al.(2015)Ronneberger, Fischer, and
  Brox]{ronneberger2015u}
Olaf Ronneberger, Philipp Fischer, and Thomas Brox.
\newblock U-net: Convolutional networks for biomedical image segmentation.
\newblock In \emph{International Conference on Medical Image Computing and
  Computer-Assisted Intervention}, pp.\  234--241. Springer, 2015.

\bibitem[Roth et~al.(2015)Roth, Lu, Farag, Shin, Liu, Turkbey, and
  Summers]{roth2015deeporgan}
Holger~R Roth, Le~Lu, Amal Farag, Hoo-Chang Shin, Jiamin Liu, Evrim~B Turkbey,
  and Ronald~M Summers.
\newblock Deeporgan: Multi-level deep convolutional networks for automated
  pancreas segmentation.
\newblock In \emph{International conference on medical image computing and
  computer-assisted intervention}, pp.\  556--564. Springer, 2015.

\bibitem[Saenz et~al.(2024)Saenz, Chen, Marklund, and
  Rajpurkar]{saenz2024maida}
Agustina Saenz, Emma Chen, Henrik Marklund, and Pranav Rajpurkar.
\newblock The maida initiative: establishing a framework for global
  medical-imaging data sharing.
\newblock \emph{The Lancet Digital Health}, 6\penalty0 (1):\penalty0 e6--e8,
  2024.

\bibitem[Sellergren et~al.(2022)Sellergren, Chen, Nabulsi, Li, Maschinot,
  Sarna, Huang, Lau, Kalidindi, Etemadi, et~al.]{sellergren2022simplified}
Andrew~B Sellergren, Christina Chen, Zaid Nabulsi, Yuanzhen Li, Aaron
  Maschinot, Aaron Sarna, Jenny Huang, Charles Lau, Sreenivasa~Raju Kalidindi,
  Mozziyar Etemadi, et~al.
\newblock Simplified transfer learning for chest radiography models using less
  data.
\newblock \emph{Radiology}, 305\penalty0 (2):\penalty0 454--465, 2022.

\bibitem[Shekoofeh et~al.(2021)Shekoofeh, Basil, Fiona, Zachary, Jan, Jonathan,
  Aaron, Alan, Simon, Ting, et~al.]{shekoofeh2021big}
Azizi Shekoofeh, Mustafa Basil, Ryan Fiona, Beaver Zachary, Freyberg Jan,
  Deaton Jonathan, Loh Aaron, Karthikesalingam Alan, Kornblith Simon, Chen
  Ting, et~al.
\newblock Big self-supervised models advance medical image classification.
\newblock \emph{arXiv preprint arXiv:2101.05224}, 2021.

\bibitem[Shin et~al.(2016)Shin, Roth, Gao, Lu, Xu, Nogues, Yao, Mollura, and
  Summers]{shin2016deep}
Hoo-Chang Shin, Holger~R Roth, Mingchen Gao, Le~Lu, Ziyue Xu, Isabella Nogues,
  Jianhua Yao, Daniel Mollura, and Ronald~M Summers.
\newblock Deep convolutional neural networks for computer-aided detection:
  {CNN} architectures, dataset characteristics and transfer learning.
\newblock \emph{IEEE transactions on medical imaging}, 35\penalty0
  (5):\penalty0 1285--1298, 2016.

\bibitem[Siddique et~al.(2021)Siddique, Paheding, Elkin, and
  Devabhaktuni]{siddique2021u}
Nahian Siddique, Sidike Paheding, Colin~P Elkin, and Vijay Devabhaktuni.
\newblock U-net and its variants for medical image segmentation: A review of
  theory and applications.
\newblock \emph{Ieee Access}, 9:\penalty0 82031--82057, 2021.

\bibitem[Steiner et~al.(2021)Steiner, Kolesnikov, Zhai, Wightman, Uszkoreit,
  and Beyer]{steiner2021train}
Andreas Steiner, Alexander Kolesnikov, Xiaohua Zhai, Ross Wightman, Jakob
  Uszkoreit, and Lucas Beyer.
\newblock How to train your vit? data, augmentation, and regularization in
  vision transformers.
\newblock \emph{arXiv preprint arXiv:2106.10270}, 2021.

\bibitem[Sun et~al.(2017)Sun, Shrivastava, Singh, and Gupta]{sun2017revisiting}
Chen Sun, Abhinav Shrivastava, Saurabh Singh, and Abhinav Gupta.
\newblock Revisiting unreasonable effectiveness of data in deep learning era.
\newblock In \emph{Proceedings of the IEEE international conference on computer
  vision}, pp.\  843--852, 2017.

\bibitem[Tajbakhsh et~al.(2016)Tajbakhsh, Shin, Gurudu, Hurst, Kendall, Gotway,
  and Liang]{tajbakhsh2016convolutional}
Nima Tajbakhsh, Jae~Y Shin, Suryakanth~R Gurudu, R~Todd Hurst, Christopher~B
  Kendall, Michael~B Gotway, and Jianming Liang.
\newblock Convolutional neural networks for medical image analysis: Full
  training or fine tuning?
\newblock \emph{IEEE transactions on medical imaging}, 35\penalty0
  (5):\penalty0 1299--1312, 2016.

\bibitem[Tang et~al.(2022)Tang, Yang, Li, Roth, Landman, Xu, Nath, and
  Hatamizadeh]{tang2022self}
Yucheng Tang, Dong Yang, Wenqi Li, Holger~R Roth, Bennett Landman, Daguang Xu,
  Vishwesh Nath, and Ali Hatamizadeh.
\newblock Self-supervised pre-training of swin transformers for 3d medical
  image analysis.
\newblock In \emph{Proceedings of the IEEE/CVF Conference on Computer Vision
  and Pattern Recognition}, pp.\  20730--20740, 2022.

\bibitem[Tao et~al.(2020)Tao, Li, Zhou, Ma, and Zheng]{tao2020revisiting}
Xing Tao, Yuexiang Li, Wenhui Zhou, Kai Ma, and Yefeng Zheng.
\newblock Revisiting rubik’s cube: Self-supervised learning with volume-wise
  transformation for 3d medical image segmentation.
\newblock In \emph{International Conference on Medical Image Computing and
  Computer-Assisted Intervention}, pp.\  238--248. Springer, 2020.

\bibitem[Team(2011)]{national2011national}
National Lung Screening Trial~Research Team.
\newblock The national lung screening trial: overview and study design.
\newblock \emph{Radiology}, 258\penalty0 (1):\penalty0 243--253, 2011.

\bibitem[Tendle \& Hasan(2021)Tendle and Hasan]{tendle2021study}
Atharva Tendle and Mohammad~Rashedul Hasan.
\newblock A study of the generalizability of self-supervised representations.
\newblock \emph{Machine Learning with Applications}, 6:\penalty0 100124, 2021.

\bibitem[Valanarasu et~al.(2023)Valanarasu, Tang, Yang, Xu, Zhao, Li, Patel,
  Landman, Xu, He, et~al.]{valanarasu2023disruptive}
Jeya Maria~Jose Valanarasu, Yucheng Tang, Dong Yang, Ziyue Xu, Can Zhao, Wenqi
  Li, Vishal~M Patel, Bennett Landman, Daguang Xu, Yufan He, et~al.
\newblock Disruptive autoencoders: Leveraging low-level features for 3d medical
  image pre-training.
\newblock \emph{arXiv preprint arXiv:2307.16896}, 2023.

\bibitem[Valindria et~al.(2018)Valindria, Pawlowski, Rajchl, Lavdas, Aboagye,
  Rockall, Rueckert, and Glocker]{valindria2018multi}
Vanya~V Valindria, Nick Pawlowski, Martin Rajchl, Ioannis Lavdas, Eric~O
  Aboagye, Andrea~G Rockall, Daniel Rueckert, and Ben Glocker.
\newblock Multi-modal learning from unpaired images: Application to multi-organ
  segmentation in ct and mri.
\newblock In \emph{2018 IEEE winter conference on applications of computer
  vision (WACV)}, pp.\  547--556. IEEE, 2018.

\bibitem[Wang et~al.(2019)Wang, Zhou, Shen, Park, Fishman, and
  Yuille]{wang2019abdominal}
Yan Wang, Yuyin Zhou, Wei Shen, Seyoun Park, Elliot~K Fishman, and Alan~L
  Yuille.
\newblock Abdominal multi-organ segmentation with organ-attention networks and
  statistical fusion.
\newblock \emph{Medical image analysis}, 55:\penalty0 88--102, 2019.

\bibitem[Wang et~al.(2022)Wang, Yu, Rao, Zhou, and Lu]{wang2022p2p}
Ziyi Wang, Xumin Yu, Yongming Rao, Jie Zhou, and Jiwen Lu.
\newblock P2p: Tuning pre-trained image models for point cloud analysis with
  point-to-pixel prompting.
\newblock \emph{Advances in neural information processing systems},
  35:\penalty0 14388--14402, 2022.

\bibitem[Wasserthal et~al.(2022)Wasserthal, Meyer, Breit, Cyriac, Yang, and
  Segeroth]{wasserthal2022totalsegmentator}
Jakob Wasserthal, Manfred Meyer, Hanns-Christian Breit, Joshy Cyriac, Shan
  Yang, and Martin Segeroth.
\newblock Totalsegmentator: robust segmentation of 104 anatomical structures in
  ct images.
\newblock \emph{arXiv preprint arXiv:2208.05868}, 2022.

\bibitem[Wei et~al.(2022)Wei, Fan, Xie, Wu, Yuille, and
  Feichtenhofer]{wei2022masked}
Chen Wei, Haoqi Fan, Saining Xie, Chao-Yuan Wu, Alan Yuille, and Christoph
  Feichtenhofer.
\newblock Masked feature prediction for self-supervised visual pre-training.
\newblock In \emph{Proceedings of the IEEE/CVF Conference on Computer Vision
  and Pattern Recognition}, pp.\  14668--14678, 2022.

\bibitem[Xia et~al.(2022)Xia, Yu, Chu, Kawamoto, Park, Liu, Chen, Zhu, Li,
  Zhou, et~al.]{xia2022felix}
Yingda Xia, Qihang Yu, Linda Chu, Satomi Kawamoto, Seyoun Park, Fengze Liu,
  Jieneng Chen, Zhuotun Zhu, Bowen Li, Zongwei Zhou, et~al.
\newblock The felix project: Deep networks to detect pancreatic neoplasms.
\newblock \emph{medRxiv}, 2022.

\bibitem[Xiao et~al.(2022)Xiao, Bai, Yuille, and Zhou]{xiao2022delving}
Junfei Xiao, Yutong Bai, Alan Yuille, and Zongwei Zhou.
\newblock Delving into masked autoencoders for multi-label thorax disease
  classification.
\newblock \emph{IEEE Winter Conference on Applications of Computer Vision},
  2022.
\newblock URL \url{https://github.com/lambert-x/medical_mae}.

\bibitem[Xie et~al.(2020)Xie, Zhang, Liao, Xia, and Shen]{xie2020pgl}
Yutong Xie, Jianpeng Zhang, Zehui Liao, Yong Xia, and Chunhua Shen.
\newblock Pgl: Prior-guided local self-supervised learning for 3d medical image
  segmentation.
\newblock \emph{arXiv preprint arXiv:2011.12640}, 2020.

\bibitem[Xie et~al.(2022)Xie, Zhang, Xia, and Wu]{xie2022unimiss}
Yutong Xie, Jianpeng Zhang, Yong Xia, and Qi~Wu.
\newblock Unimiss: Universal medical self-supervised learning via breaking
  dimensionality barrier.
\newblock In \emph{European Conference on Computer Vision}, pp.\  558--575.
  Springer, 2022.

\bibitem[Yang et~al.(2020)Yang, He, Liang, Yang, Zhang, and
  Xie]{yang2020transfer}
Xingyi Yang, Xuehai He, Yuxiao Liang, Yue Yang, Shanghang Zhang, and Pengtao
  Xie.
\newblock Transfer learning or self-supervised learning? a tale of two
  pretraining paradigms.
\newblock \emph{arXiv preprint arXiv:2007.04234}, 2020.

\bibitem[Yosinski et~al.(2014)Yosinski, Clune, Bengio, and
  Lipson]{yosinski2014transferable}
Jason Yosinski, Jeff Clune, Yoshua Bengio, and Hod Lipson.
\newblock How transferable are features in deep neural networks?
\newblock In \emph{Advances in neural information processing systems}, pp.\
  3320--3328, 2014.

\bibitem[You et~al.(2022)You, Zhao, Liu, Dong, Chinchali, Topcu, Staib, and
  Duncan]{you2022class}
Chenyu You, Ruihan Zhao, Fenglin Liu, Siyuan Dong, Sandeep Chinchali, Ufuk
  Topcu, Lawrence Staib, and James Duncan.
\newblock Class-aware adversarial transformers for medical image segmentation.
\newblock \emph{Advances in Neural Information Processing Systems},
  35:\penalty0 29582--29596, 2022.

\bibitem[Zamir et~al.(2018)Zamir, Sax, Shen, Guibas, Malik, and
  Savarese]{zamir2018taskonomy}
Amir~R Zamir, Alexander Sax, William Shen, Leonidas~J Guibas, Jitendra Malik,
  and Silvio Savarese.
\newblock Taskonomy: Disentangling task transfer learning.
\newblock In \emph{Proceedings of the IEEE conference on computer vision and
  pattern recognition}, pp.\  3712--3722, 2018.

\bibitem[Zhai et~al.(2022)Zhai, Kolesnikov, Houlsby, and
  Beyer]{zhai2022scaling}
Xiaohua Zhai, Alexander Kolesnikov, Neil Houlsby, and Lucas Beyer.
\newblock Scaling vision transformers.
\newblock In \emph{Proceedings of the IEEE/CVF Conference on Computer Vision
  and Pattern Recognition}, pp.\  12104--12113, 2022.

\bibitem[Zhang et~al.(2021)Zhang, Xie, Xia, and Shen]{zhang2021dodnet}
Jianpeng Zhang, Yutong Xie, Yong Xia, and Chunhua Shen.
\newblock Dodnet: Learning to segment multi-organ and tumors from multiple
  partially labeled datasets.
\newblock In \emph{Proceedings of the IEEE/CVF Conference on Computer Vision
  and Pattern Recognition}, pp.\  1195--1204, 2021.

\bibitem[Zhang \& Metaxas(2023)Zhang and Metaxas]{zhang2023challenges}
Shaoting Zhang and Dimitris Metaxas.
\newblock On the challenges and perspectives of foundation models for medical
  image analysis.
\newblock \emph{arXiv preprint arXiv:2306.05705}, 2023.

\bibitem[Zhao et~al.(2023)Zhao, Zhang, Wu, Zhang, Zhang, Wang, and
  Xie]{zhao2023one}
Ziheng Zhao, Yao Zhang, Chaoyi Wu, Xiaoman Zhang, Ya~Zhang, Yanfeng Wang, and
  Weidi Xie.
\newblock One model to rule them all: Towards universal segmentation for
  medical images with text prompts.
\newblock \emph{arXiv preprint arXiv:2312.17183}, 2023.

\bibitem[Zhou et~al.(2021{\natexlab{a}})Zhou, Wei, Wang, Shen, Xie, Yuille, and
  Kong]{zhou2021ibot}
Jinghao Zhou, Chen Wei, Huiyu Wang, Wei Shen, Cihang Xie, Alan Yuille, and Tao
  Kong.
\newblock ibot: Image bert pre-training with online tokenizer.
\newblock \emph{arXiv preprint arXiv:2111.07832}, 2021{\natexlab{a}}.

\bibitem[Zhou et~al.(2022{\natexlab{a}})Zhou, Liu, Qiao, Xiang, and
  Loy]{zhou2022domain}
Kaiyang Zhou, Ziwei Liu, Yu~Qiao, Tao Xiang, and Chen~Change Loy.
\newblock Domain generalization: A survey.
\newblock \emph{IEEE Transactions on Pattern Analysis and Machine
  Intelligence}, 2022{\natexlab{a}}.

\bibitem[Zhou(2021)]{zhou2021towards}
Zongwei Zhou.
\newblock \emph{Towards Annotation-Efficient Deep Learning for Computer-Aided
  Diagnosis}.
\newblock PhD thesis, Arizona State University, 2021.
\newblock URL \url{https://github.com/MrGiovanni/Dissertation}.

\bibitem[Zhou et~al.(2017)Zhou, Shin, Zhang, Gurudu, Gotway, and
  Liang]{zhou2017fine}
Zongwei Zhou, Jae Shin, Lei Zhang, Suryakanth Gurudu, Michael Gotway, and
  Jianming Liang.
\newblock Fine-tuning convolutional neural networks for biomedical image
  analysis: actively and incrementally.
\newblock In \emph{IEEE/CVF Conference on Computer Vision and Pattern
  Recognition}, pp.\  7340--7351, 2017.
\newblock URL \url{https://github.com/MrGiovanni/Active-Learning}.

\bibitem[Zhou et~al.(2018)Zhou, Siddiquee, Tajbakhsh, and
  Liang]{zhou2018unet++}
Zongwei Zhou, Md~Mahfuzur~Rahman Siddiquee, Nima Tajbakhsh, and Jianming Liang.
\newblock Unet++: A nested u-net architecture for medical image segmentation.
\newblock In \emph{Deep Learning in Medical Image Analysis and Multimodal
  Learning for Clinical Decision Support}, pp.\  3--11. Springer, 2018.
\newblock URL \url{https://github.com/MrGiovanni/UNetPlusPlus}.

\bibitem[Zhou et~al.(2019{\natexlab{a}})Zhou, Siddiquee, Tajbakhsh, and
  Liang]{zhou2019unet++}
Zongwei Zhou, Md~Mahfuzur~Rahman Siddiquee, Nima Tajbakhsh, and Jianming Liang.
\newblock Unet++: Redesigning skip connections to exploit multiscale features
  in image segmentation.
\newblock \emph{IEEE Transactions on Medical Imaging}, 39\penalty0
  (6):\penalty0 1856--1867, 2019{\natexlab{a}}.
\newblock URL \url{https://github.com/MrGiovanni/UNetPlusPlus}.

\bibitem[Zhou et~al.(2019{\natexlab{b}})Zhou, Sodha, Siddiquee, Feng,
  Tajbakhsh, Gotway, and Liang]{zhou2019models}
Zongwei Zhou, Vatsal Sodha, Md~Mahfuzur~Rahman Siddiquee, Ruibin Feng, Nima
  Tajbakhsh, Michael~B Gotway, and Jianming Liang.
\newblock Models genesis: Generic autodidactic models for 3d medical image
  analysis.
\newblock In \emph{International Conference on Medical Image Computing and
  Computer-Assisted Intervention}, pp.\  384--393. Springer,
  2019{\natexlab{b}}.
\newblock URL \url{https://github.com/MrGiovanni/ModelsGenesis}.

\bibitem[Zhou et~al.(2021{\natexlab{b}})Zhou, Sodha, Pang, Gotway, and
  Liang]{zhou2021models}
Zongwei Zhou, Vatsal Sodha, Jiaxuan Pang, Michael~B Gotway, and Jianming Liang.
\newblock Models genesis.
\newblock \emph{Medical Image Analysis}, 67:\penalty0 101840,
  2021{\natexlab{b}}.
\newblock URL \url{https://github.com/MrGiovanni/ModelsGenesis}.

\bibitem[Zhou et~al.(2022{\natexlab{b}})Zhou, Gotway, and
  Liang]{zhou2022interpreting}
Zongwei Zhou, Michael~B Gotway, and Jianming Liang.
\newblock Interpreting medical images.
\newblock In \emph{Intelligent Systems in Medicine and Health}, pp.\  343--371.
  Springer, 2022{\natexlab{b}}.

\bibitem[Zoph et~al.(2020)Zoph, Ghiasi, Lin, Cui, Liu, Cubuk, and
  Le]{zoph2020rethinking}
Barret Zoph, Golnaz Ghiasi, Tsung-Yi Lin, Yin Cui, Hanxiao Liu, Ekin~Dogus
  Cubuk, and Quoc Le.
\newblock Rethinking pre-training and self-training.
\newblock \emph{Advances in neural information processing systems},
  33:\penalty0 3833--3845, 2020.

\end{thebibliography}
\bibliographystyle{iclr2024_conference}

\newpage
\appendix

\renewcommand \thepart{}
\renewcommand \partname{}

\part{Appendix} 
\setcounter{secnumdepth}{4}
\setcounter{tocdepth}{4}
\parttoc 

\clearpage
\section{An Extensive Dataset: \ourdataset}\label{sec:dataset_appendix}
\begin{table*}[h]
    \caption{\textbf{An extensive dataset of \totalct\ CT volumes with per-voxel annotations of 25 anatomical structures.} \ourdataset\ marks a breakthrough in data and annotation scales, encompassing \totalmasks\ organ and tumor masks and \totalslices\ annotated images sourced from 88 global hospitals in 19 countries. In 2009, prior to the creation of ImageNet \citep{deng2009imagenet}, AI models struggled with general image representation due to limited data availability, a challenge still prevalent in 3D medical image analysis today. As demonstrated in the table, existing public datasets often suffer from limited, partial, and incomplete annotations, and exhibit biases towards certain populations, medical centers, and countries. Our dataset not only addresses these shortcomings but has also been refined to eliminate redundancy, detailing the count of unique CT volumes sourced from each existing dataset incorporated into ours. \ourdataset\ offers a diverse and extensive range of annotated data, thus marking a significant advancement in the field.} 
    \centering
    \scriptsize
    \begin{tabular}{p{0.35\linewidth}P{0.05\linewidth}P{0.06\linewidth}P{0.05\linewidth}P{0.13\linewidth}P{0.17\linewidth}}
    \toprule
    dataset (year) [source] & \makecell{\# of\\organ} & \makecell{\# of\\volume} & \makecell{\# of\\center} & \makecell{source\\countries} & license \\
    \midrule
    
    \rowcolor{igray!4}1. Pancreas-CT \citeyearpar{roth2015deeporgan} [\href{https://academictorrents.com/details/80ecfefcabede760cdbdf63e38986501f7becd49}{link}] & 1 & 82 & 1 & US & CC BY 3.0 \\

    2. CHAOS \citeyearpar{valindria2018multi} [\href{https://chaos.grand-challenge.org/Download/}{link}] & 4 & 40 & 1 & TR & CC BY-SA 4.0 \\

    \rowcolor{igray!4}3. CT-ORG \citeyearpar{rister2020ct} [\href{https://wiki.cancerimagingarchive.net/pages/viewpage.action?pageId=61080890#61080890cd4d3499fa294f489bf1ea261184fd24}{link}] & 5 & 140 & 8 & \makecell{DE, NL, CA,\\FR, IL, US} & CC BY 3.0\\ 

    4. BTCV \citeyearpar{landman2015miccai} [\href{https://www.synapse.org/#!Synapse:syn3193805/wiki/89480}{link}] & 12 & 50 & 1 & US & CC BY 4.0 \\

    \rowcolor{igray!4}5. AMOS22 \citeyearpar{ji2022amos} [\href{https://amos22.grand-challenge.org}{link}] & 15 & 500 & 2 & CN & CC BY-NC-SA \\

    6. WORD \citeyearpar{luo2021word} [\href{https://github.com/HiLab-git/WORD}{link}] & 16 & 150 & 1 & CN & GNU GPL 3.0 \\

    \rowcolor{igray!4}7-12. MSD CT Tasks \citeyearpar{antonelli2021medical} [\href{https://decathlon-10.grand-challenge.org/}{link}] & 9 & 947 & 1 & US & CC BY-SA 4.0 \\

    13. LiTS \citeyearpar{bilic2019liver} [\href{https://competitions.codalab.org/competitions/17094}{link}] & 1 & 201 & 7 & \makecell{DE, NL, CA,\\FR, IL} & CC BY-SA 4.0 \\

    \rowcolor{igray!4}14. AbdomenCT-1K \citeyearpar{ma2021abdomenct} [\href{https://github.com/JunMa11/AbdomenCT-1K}{link}] & 4 & 1,050 & 12 & \makecell{DE, NL, CA, FR, \\IL, US, CN} & CC BY-NC-SA \\

    15. KiTS \citeyearpar{heller2020international} [\href{https://kits-challenge.org/kits23/}{link}] & 1 & 300 & 1 & US & CC BY-NC-SA 4.0 \\

    \rowcolor{igray!4}16. FLARE'23 \citeyearpar{ma2022fast} [\href{https://codalab.lisn.upsaclay.fr/competitions/12239}{link}] & 13 & 4,000 & 30 & - & CC BY-NC-ND 4.0 \\

    17. Trauma Det.~\citeyearpar{rsna-2023-abdominal-trauma-detection} [\href{https://www.rsna.org/education/ai-resources-and-training/ai-image-challenge/abdominal-trauma-detection-ai-challenge}{link}] & 0 & 4,711 & 23 & \makecell{CL, DE, ES, TR, \\AUS, TH, TW, \\MA, MT, CA, \\IE, BR, BA} & - \\
    
    \midrule

    \rowcolor{igray!4}18. FUMPE \citeyearpar{masoudi2018new} [\href{https://figshare.com/collections/FUMPE/4107803/1}{link}] & 1 & 35 & 1 & IR & CC BY 4.0 \\

    19. KiPA22~\citeyearpar{he2021meta} [\href{https://kipa22.grand-challenge.org}{link}] & 4 & 100 & 1 & CN & CC BY-NC-ND 3.0\\
    
    \rowcolor{igray!4}20. AATTCT-IDS~\citeyearpar{ma2023aatct} [\href{https://figshare.com/articles/dataset/AATTCT-IDS/23807256}{link}] & 0 & 300 & 1 & CN & - \\

    21. CTSpine1K~\citeyearpar{deng2021ctspine1k} [\href{https://paperswithcode.com/dataset/ctspine1k}{link}] & 26 & 1,005 & - & - & CC BY 4.0 \\
       
    \rowcolor{igray!4}22. AutoPET~\citeyearpar{gatidis2022whole} [\href{https://autopet.grand-challenge.org/Description/}{link}] & 0 & 1,014 & 2 & DE & TCIA Restricted \\

    23. TotalSegmentator~\citeyearpar{wasserthal2022totalsegmentator} [\href{https://doi.org/10.5281/zenodo.6802613}{link}] & 104 & 1,204 & 1 & CH & CC BY 4.0 \\
    
    \midrule
    \rowcolor{igray!4}24. AbdomenAtlas 1.0~\citeyearpar{qu2023annotating} [\href{https://huggingface.co/datasets/AbdomenAtlas/AbdomenAtlas_1.0_Mini}{link}] & 9 & 5,195 & 26 & \makecell{US, DE, NL, CA\\FR, IL, CN, TR } & CC BY-NC-SA 4.0 \\
    
    25. \ourdataset & 25 & \totalct\ & 88 & \makecell{MT, IE, BR, BA, \\AUS, TH, TW, \\CA, TR, CH, \\CL, ES, MA, \\US, DE, NL,\\FR, IL, CN } & CC BY-NC-SA 4.0 \\
    \bottomrule
    \end{tabular}
    \begin{tablenotes}
        \item US: United States \quad DE: Germany \quad NL: Netherlands \quad CA: Canada \quad FR: France \quad IL: Israel \quad IR: Iran
        \item CN: China \quad TR: Turkey \quad CH: Switzerland \quad AUS: Australia \quad TH: Thailand \quad TW: Taiwan \quad CL: Chile
        \item ES: Spain \quad MA: Morocco \quad MT: Malta \quad IE: Ireland \quad BR: Brazil \quad BA: Bosnia and Herzegowina
    \end{tablenotes}
    \label{tab:dataset_overview_appendix}
\end{table*}

\begin{figure*}[t]
    \centering
    \includegraphics[width=1.0\linewidth]{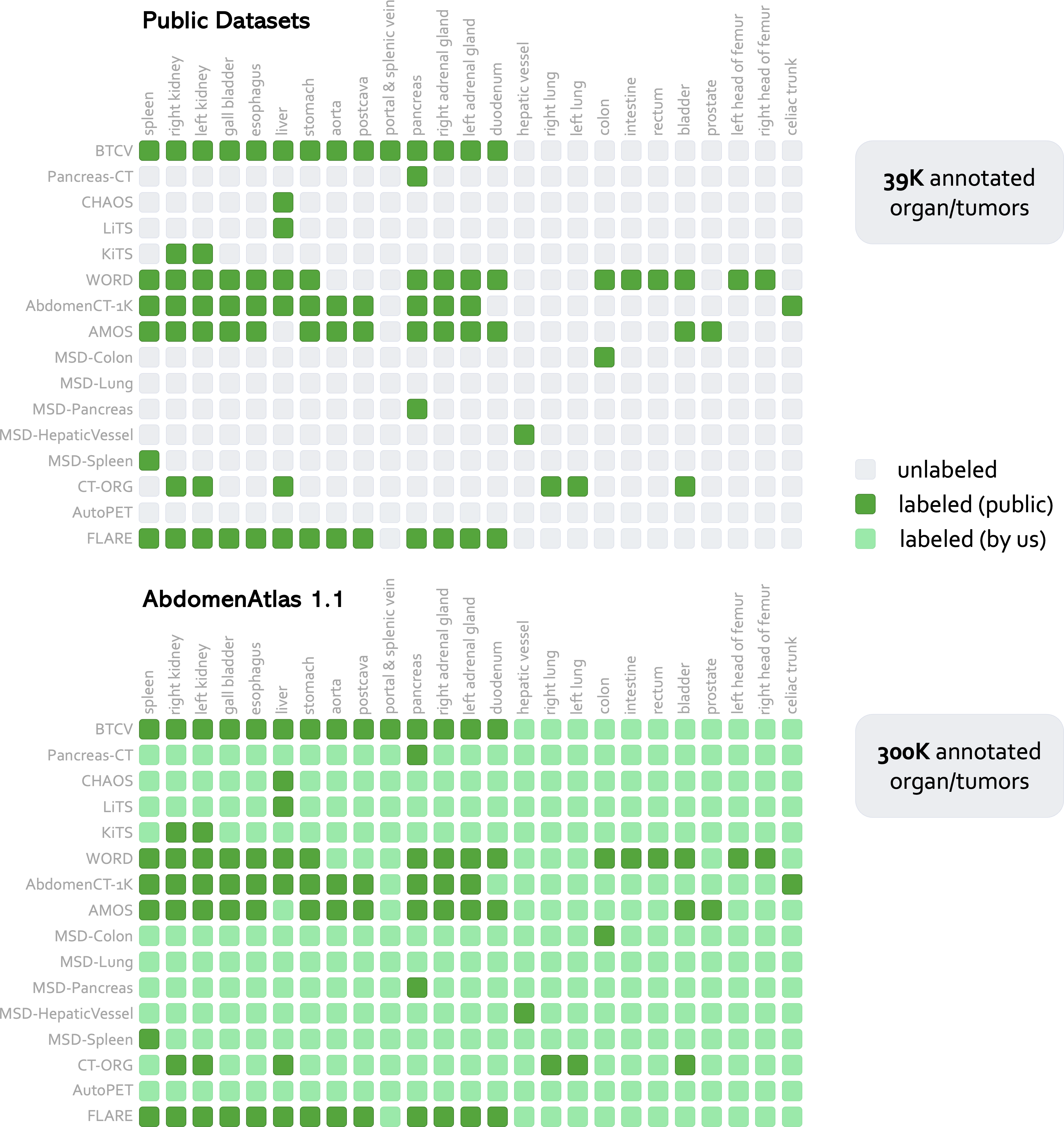}
    \caption{
    \textbf{Evolution from a combination of public data to \ourdataset.} \ourdataset\ is NOT a simple combination of existing datasets. The \totalct\ CT volumes in the combination of public datasets only contain a total of \textbf{39K} annotated organ masks while our \ourdataset\ provides over \textbf{\totalmasks} annotated organ/tumor masks for these CT volumes, substantially increasing the number of masks by \textbf{6.4} times.
    Creating \totalmasks\ high-quality organ/tumor masks for \totalct\ CT volumes requires extensive medical knowledge and annotation costs (much more difficult than annotating natural images). Based on our experience and those reported in \citet{park2020annotated}, trained radiologists annotate abdominal organs at a rate of 30--60 minutes per organ per CT volume. This translates to \textbf{247K} human hours for completing \ourdataset. We employed a highly efficient annotation method, combining AI with the expertise of ten radiologists using active learning (details in \appendixname\ \ref{sec:annotation_standard_appendix}), to overcome this challenge and produce the largest annotated dataset to date.
    }
    \label{fig:dataset_completion}
\end{figure*}

\clearpage
\subsection{Domain Transfer Across Datasets}\label{sec:domain_gaps_appendix}

\begin{figure*}[h]
    \centering
    \includegraphics[width=1.0\linewidth]{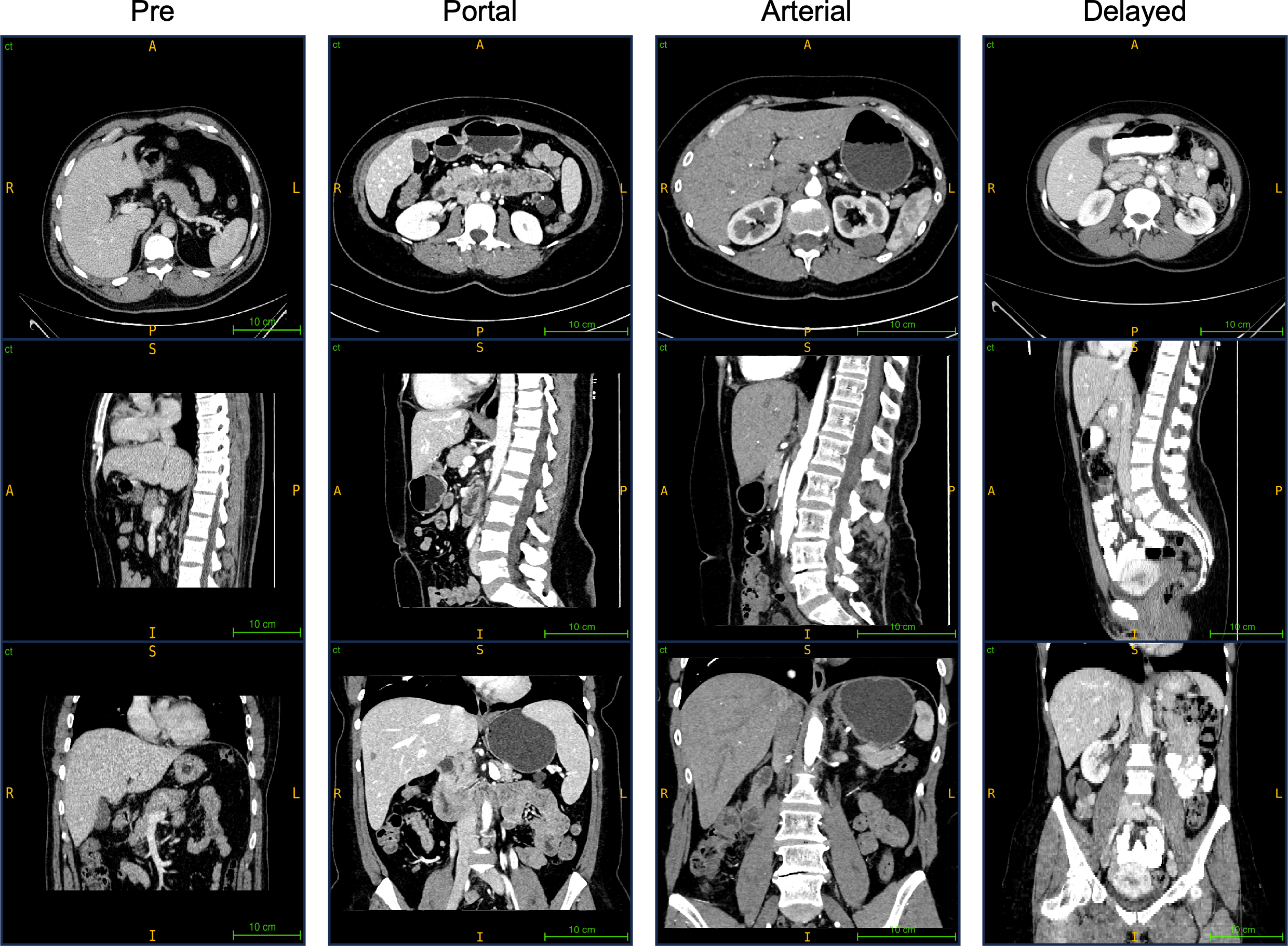}
    \caption{
    \textbf{Domain gaps.} Examples of CT volumes from different domains (e.g., hospitals and countries) illustrate the variability in images.
    \ourdataset\ are created by a large variety of CT scanners, imaging protocols, and acquired from numerous hospitals worldwide (\tableautorefname~\ref{tab:dataset_overview}). We note that substantial differences in CT volumes occur in image quality and technical display, originating from different acquisition parameters, reconstruction kernels, and contrast enhancements. 
    }
    \label{fig:domain_gaps_appendix}
\end{figure*}

\tableautorefname~\ref{tab:external_datasets} shows that \ourmodel\ is pretty robust because our \ourdataset\ covers a variety of domains (i.e., 88 hospitals with different scanners and protocols), as shown in \figureautorefname~\ref{fig:domain_gaps_appendix}; models pre-trained on this dataset are expected to be generalizable for novel domains. Therefore, domain transfer becomes less important if the model is pre-trained on large and diverse datasets, elaborating on the two points below.

\begin{enumerate}
    
    \item The domain transfer problem could be solved by methodology innovation, and also by training AI models on enormous datasets. This point has been more clear recently demonstrated by large language models (GPT) and vision foundation models (SAM), which show incredible performance in the ``novel domain''. However, this achievement may not be directly attributed to method-driven solutions for domain transfer, but simply because the AI might have been trained on similar sentences or images. This was also pointed out by Yann Lecun---\textit{beware of testing on the training set}---in response to the incredible results achieved by GPT.

    \item In some sense, our paper explores dataset-driven solutions for domain transfer. The robust performance of our models when direct inference on multiple domains could also be attributed to our large-scale, fully-annotated medical dataset---as one of our major contributions. The release of \ourdataset\ can foster AI models that are more robust than the majority of existing models that are only trained on a few hundred CT volumes from limited domains. In addition, existing domain transfer methods could also be supplemented with direct inference and fine-tuning to further improve AI performance.
    
\end{enumerate}

\clearpage
\subsection{Uniform Annotation Standards}\label{sec:annotation_standard_appendix}

\begin{figure}[h]
\centering
\includegraphics[width=1\columnwidth]{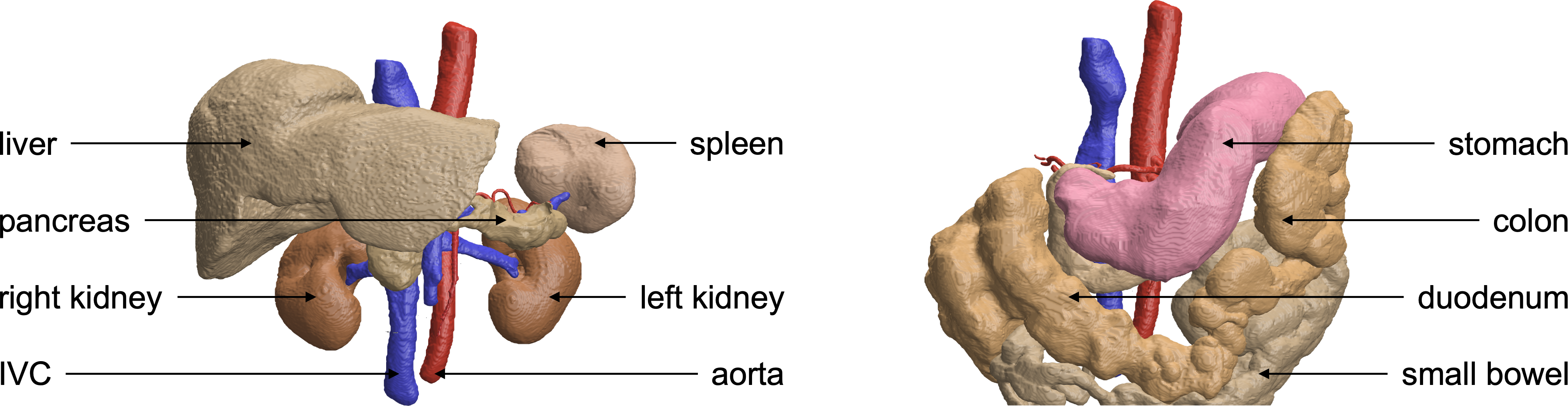}
    \caption{
    \textbf{\textit{Automated organ annotations.}} Our annotation pipeline involved an interactive segmentation approach, a synergy of AI algorithms and human expertise, which promises to improve efficiency while upholding high-quality annotations. \textit{Four senior radiologists} revised the annotations predicted by our AI models, and in turn, the AI models improved their predictions by learning from these revised annotations. This interactive procedure continued to enhance the quality of annotations until no major revision was needed. Subsequently, \textit{Six junior radiologists} examine the final visualizations for accuracy (examples of the rendered images are shown above). The junior radiologists were responsible for reviewing the correctness of the annotations and marking the patient ID for any major discrepancies. Such cases are then reviewed by senior radiologists. Our uniform annotation standards, largely overlapping with those in \citet{ma2023unleashing}, require trained radiologists to spend approximately 30–60 minutes annotating each organ in a three-dimensional CT volume. 
    }
\label{fig:organ_rendering_appendix}
\end{figure}

\textbf{\textit{Automated (pseudo) tumor annotations.}} We have established uniform annotation standards for tumors, with both senior and junior radiologists actively refining and adhering to these guidelines.

\begin{itemize}
    \item Liver tumors: Liver tumors include primary tumor lesions and metastases in the liver. Annotations should encompass the entire tumor, including any invasive parts, necrosis, hemorrhage, fibrous scars, and calcifications. Healthy areas or unrelated lesions are not included.

    \item Kidney tumors: Kidney tumors include both benign and malignant tumor lesions growing in the kidneys. The entire tumor and its invasive parts to surrounding areas, plus internal changes like necrosis and calcification, should be annotated. Exclude healthy structures.

    \item Pancreatic tumors: Pancreatic tumors include all benign and malignant tumor lesions growing in the pancreas. Annotations cover the whole tumor and its invasive growth into adjacent areas, including changes like cysts, necrosis, and calcification. Exclude healthy structures.

    \item Colon tumors: Colon tumors include all benign and malignant tumor lesions developing from the colon wall. The entire tumor and its invasion into nearby structures, along with internal changes like necrosis, should be annotated, excluding healthy areas.

    \item Hepatic vessel tumors: Hepatic vessel tumors include all primary tumor lesions developing from the intrahepatic vessel wall and tumor thrombus in intrahepatic vessels. Annotations should include the tumor within the vessels, excluding external parts and unrelated lesions.

\end{itemize}

Overall, \ourdataset\ offers 51.8K pseudo tumor masks visually inspected by radiologists, though without biopsy confirmation. While these masks lack pathological validation, we anticipate they will serve as a valuable foundation for expanding precise tumor annotations in future research.

\clearpage
\section{A Suite of Pre-trained Models: \ourmodel}\label{sec:model_appendix}

\subsection{Supervised and Self-supervised Benchmarking}\label{sec:benchmark_results_appendix}

\subsubsection{Background \& Statement}

The goal of \tableautorefname~\ref{tab:model_zoo} and Appendix~\tableautorefname~\ref{tab:model_summary_appendix} is to provide a practical benchmark for the transfer learning ability of readily available pre-trained models. Our intent is not to compare the specific pre-training methodologies of each model for two primary reasons.
\begin{enumerate}
    
    \item The majority of researchers tend to fine-tune pre-existing models rather than retrain them from scratch due to convenience and accessibility. 
    
    \item Reproducing these models would require specialized hyper-parameter tuning and varied computational resources. For example, models like Swin UNETR \citep{tang2022self} were pre-trained using large-scale GPU clusters at NVIDIA, making them challenging for us to faithfully retrain.
\end{enumerate} 

Considering both practical user scenarios and computational constraints, we decided to directly use their released models and fine-tune them with consistent settings on the same datasets.

However, using existing pre-trained models can inevitably lead to certain problems. For example, the U-Net family has seen numerous variations over the years \citep{zhou2018unet++,zhou2019unet++,siddique2021u,li2024ultra}. Pre-trained models released before 2021 typically employed a basic version of U-Net \citep{zhou2019models, chen2019med3d}. On the other hand, our U-Net benefits from a more advanced code base, thanks to the \href{https://monai.io/}{MONAI} platform at NVIDIA, which includes enhanced architectures and advanced training optimization strategies. Consequently, our U-Net, even trained from scratch, is capable of surpassing the performance of these older baseline models.

\subsubsection{Implementation details of pre-training} 

For benchmark purposes (Tables~\ref{tab:model_zoo}, \ref{tab:novel_classes} and Figures~\ref{fig:few_shot}, \ref{fig:data_compute_effciency}b, \ref{fig:pancreas_tumor_classification_task}), we pre-trained U-Net, Swin UNETR, and SegResNet on 2,100 fully annotated CT volumes with 25 anatomical structures and pseudo annotations of seven tumors. The best model was selected based on the largest average DSC over the 32 classes on 310 CT volumes as the validation set. We randomly crop sub-volumes, sized 96$\times$96$\times$96 voxels, from the original CT volumes. Our \ourmodel\ is pre-trained with AdamW using $\beta_1=0.9$ and $\beta_2=0.999$ with a batch size of 2 per GPU and a cosine learning rate schedule with a warm-up for the first 100 epochs. We start with an initial learning rate of $1e^{-4}$ and a decay of $1e^{-5}$. The pre-training has been conducted on four NVIDIA A100 using multi-GPU (4) with distributed data parallel (DDP), implemented in MONAI 0.9.0., with a maximum of 800 epochs. We use the binary cross-entropy and Dice Similarity Coefficient (DSC) losses as the objective function for pre-training.

\clearpage
\subsubsection{Implementation details of fine-tuning} 

We fine-tune the pre-trained models using TotalSegmentator and the proprietary dataset datasets. During fine-tuning, configurations from pre-training persist, but we adjust the warm-up scheduler to 20 epochs, set a maximum of 200 epochs, and use a single GPU.

\begin{table}[h]
\centering
\caption{
\textbf{Benchmarking all the self-supervised and supervised models.} All the publicly available pre-trained models can be downloaded from our \href{https://huggingface.co/MrGiovanni/SuPreM/tree/main}{model repository} at Huggingface. We will continue to include more 3D pre-trained models when they are available.
}
\centering
\scriptsize
\begin{tabular}{p{0.07\linewidth}p{0.27\linewidth}p{0.12\linewidth}p{0.07\linewidth}p{0.18\linewidth}P{0.1\linewidth} }
\toprule
& name & backbone & params & pre-trained data & performance$^{\dagger}$ \\
\midrule
\multirow{7}{*}{\makecell[l]{self-\\supervised}} 
& Models Genesis \citep{zhou2019models} & U-Net & 19.08M & 623 CT volumes & 90.1 \\
 & UniMiSS \citep{xie2022unimiss} & nnU-Net & 61.79M & 5,022 CT\&MRI volumes & 92.9 \\
\cmidrule{2-6}
 & NV$^*$ & Swin UNETR & 62.19M & 1,000 CT volumes & 93.2 \\
 & NV$^*$ & Swin UNETR & 62.19M & 3,000 CT volumes & 93.4 \\
 & NV \citep{tang2022self} & Swin UNETR & 62.19M & 5,050 CT volumes & 93.8 \\
 & NV$^*$ & Swin UNETR & 62.19M & 5,050 CT volumes & 94.2 \\
 & NV$^*$ & Swin UNETR & 62.19M & \totalct\ CT volumes & 94.3 \\
\midrule
\multirow{7}{*}{\makecell[l]{supervised}} 
& Med3D \citep{chen2019med3d} & Residual U-Net & 85.75M & 1,638 CT volumes & 91.4\\
 & DoDNet \citep{zhang2021dodnet} & U-Net & 17.29M & 920 CT volumes & 93.8\\
 & DoDNet$^*$ & U-Net & 17.29M & 920 CT volumes & 94.4 \\
 & Universal Model \citep{liu2023clip} & Swin UNETR  & 62.19M & 2,100 CT volumes & 94.1\\
\cmidrule{2-6} 
 & \ourmodel$^*$ & U-Net  & 19.08M & 2,100 CT volumes & \textbf{95.4} \\
 & \ourmodel$^*$ & Swin UNETR  & 62.19M & 2,100 CT volumes & 94.6\\
 & \ourmodel$^*$ & SegResNet  & 4.7M & 2,100 CT volumes & 94.0\\
\bottomrule
\end{tabular}
\begin{tablenotes}
    \item $^{\dagger}$We report the transfer learning performance of muscle segmentation on TotalSegmentator.
    \item $^*$The name with a star ($^*$) denotes it is implemented by us and pre-trained using our \ourdataset.
\end{tablenotes}
\label{tab:model_summary_appendix}
\end{table}

\begin{figure}[h]
\centering
\includegraphics[width=1\columnwidth]{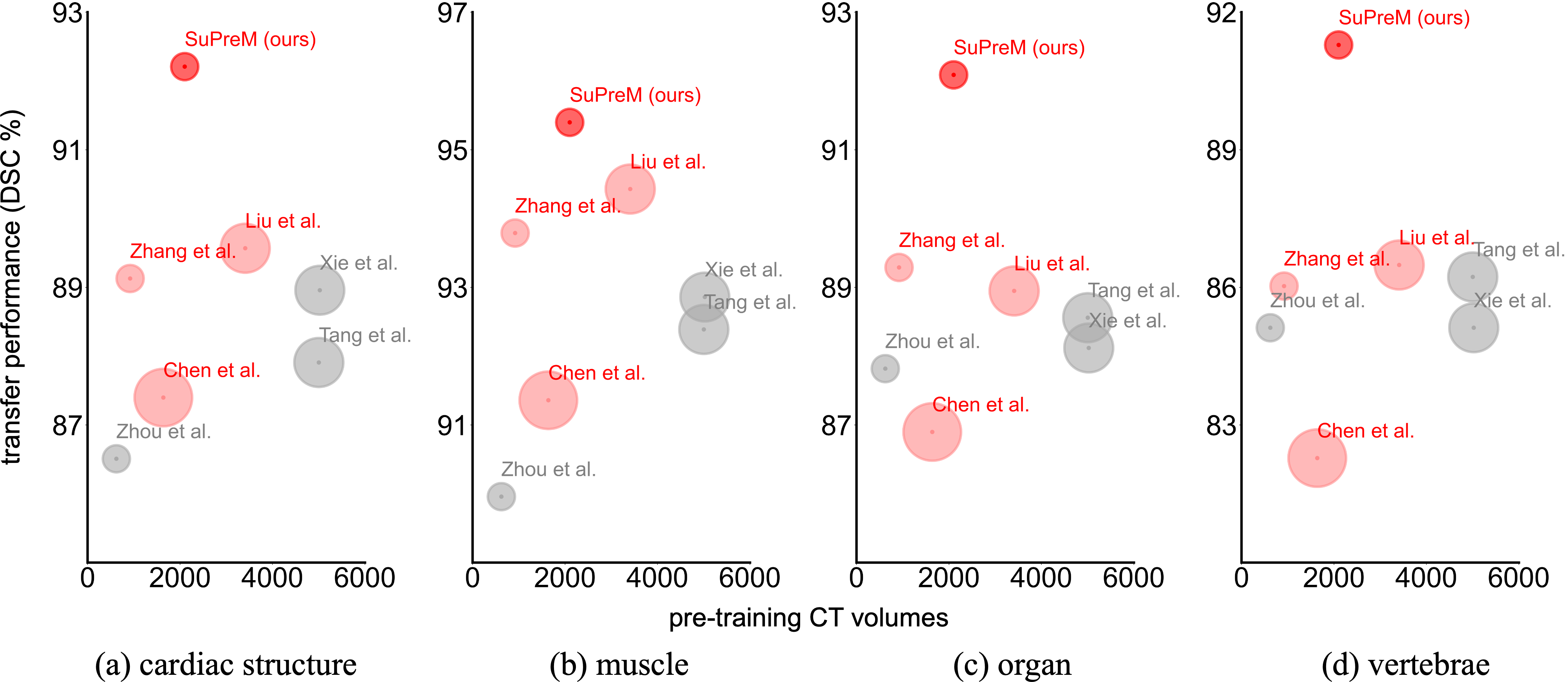}
    \caption{
    \textbf{A comprehensive benchmark on supervised and self-supervised models.} We present the segmentation performance achieved by fine-tuning models using the entire TotalSegmentator training set ($N=1,081$ annotated CT volumes) as target tasks. A larger circle size denotes a greater number of model parameters. Overall, for target tasks, supervised models (in \textcolor{red}{red}) transfer better for pre-training in comparison with self-supervised models (in \textcolor{gray}{gray}).
    }
\label{fig:comprehensive_benchmark_appendix}
\end{figure}

\clearpage

\section{Data, Annotation, and Computational Efficiency}\label{sec:efficiency_appendix}

\subsection{Annotation Efficiency in Fine-tuning}
\begin{figure*}[h]
\centering
\includegraphics[width=1.0\columnwidth]{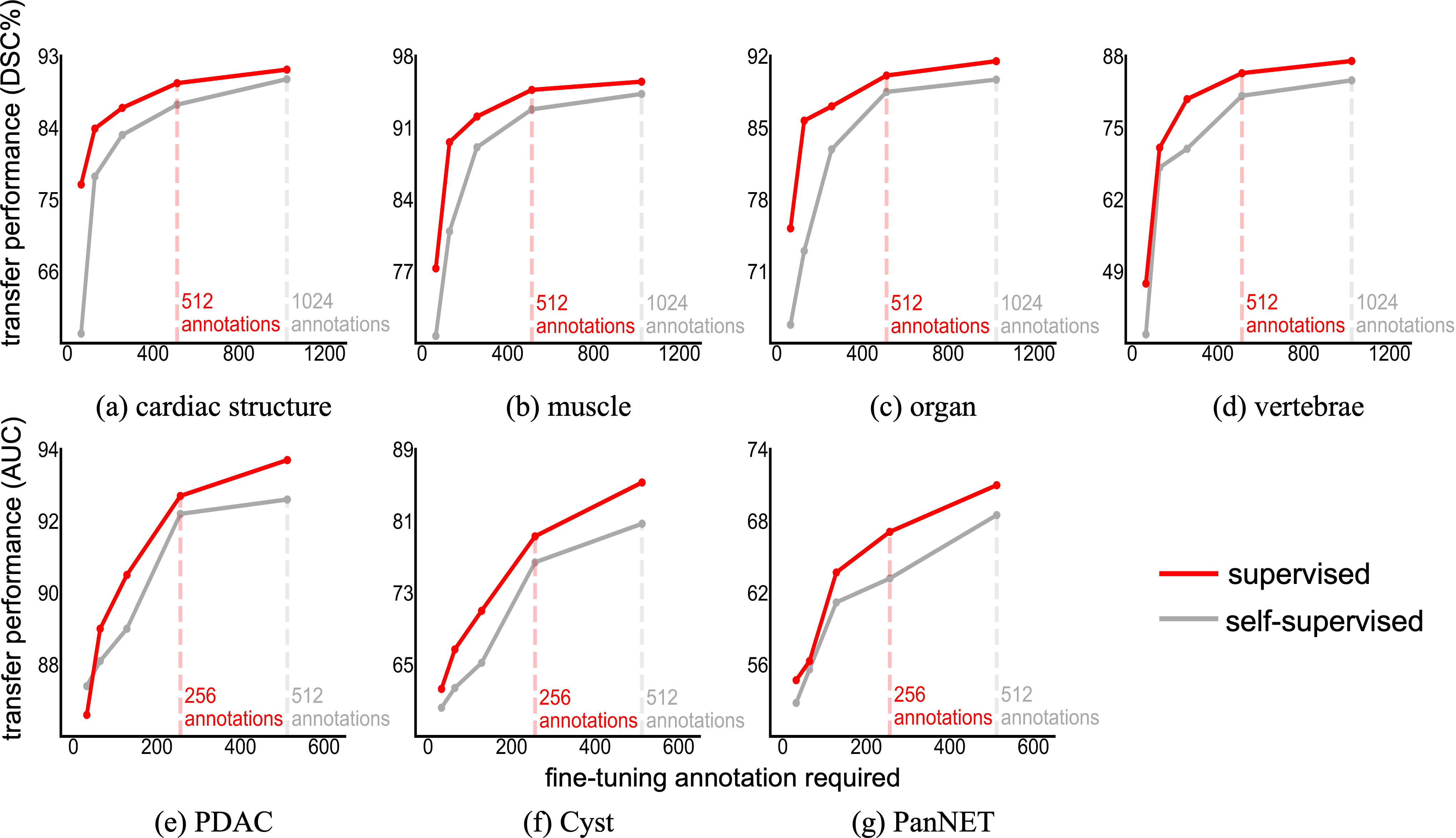}
\caption{
\textbf{\ourmodel\ is annotation efficient when transferred to novel class segmentation tasks.} We assess the annotation \& learning efficiency by fine-tuning models on different numbers of annotated CT volumes from TotalSegmentator and the proprietary dataset of a total of 66 novel classes. Specifically, TotalSegmentator provides 19 muscles, 15 cardiac structures, 5 organs, and 24 vertebrae; the proprietary dataset offers three sub-types of pancreatic tumors, including pancreatic ductal adenocarcinoma (PDAC), pancreatic cysts, and pancreatic neuroendocrine tumors (PanNET). 
}
\label{fig:all_annotation_efficiency}
\end{figure*}
\clearpage

\subsection{Convergence and Learning Efficiency in Fine-tuning}\label{sec:convergence_efficiency_appendix}

\begin{figure*}[h]
    \centering
    \includegraphics[width=\linewidth]{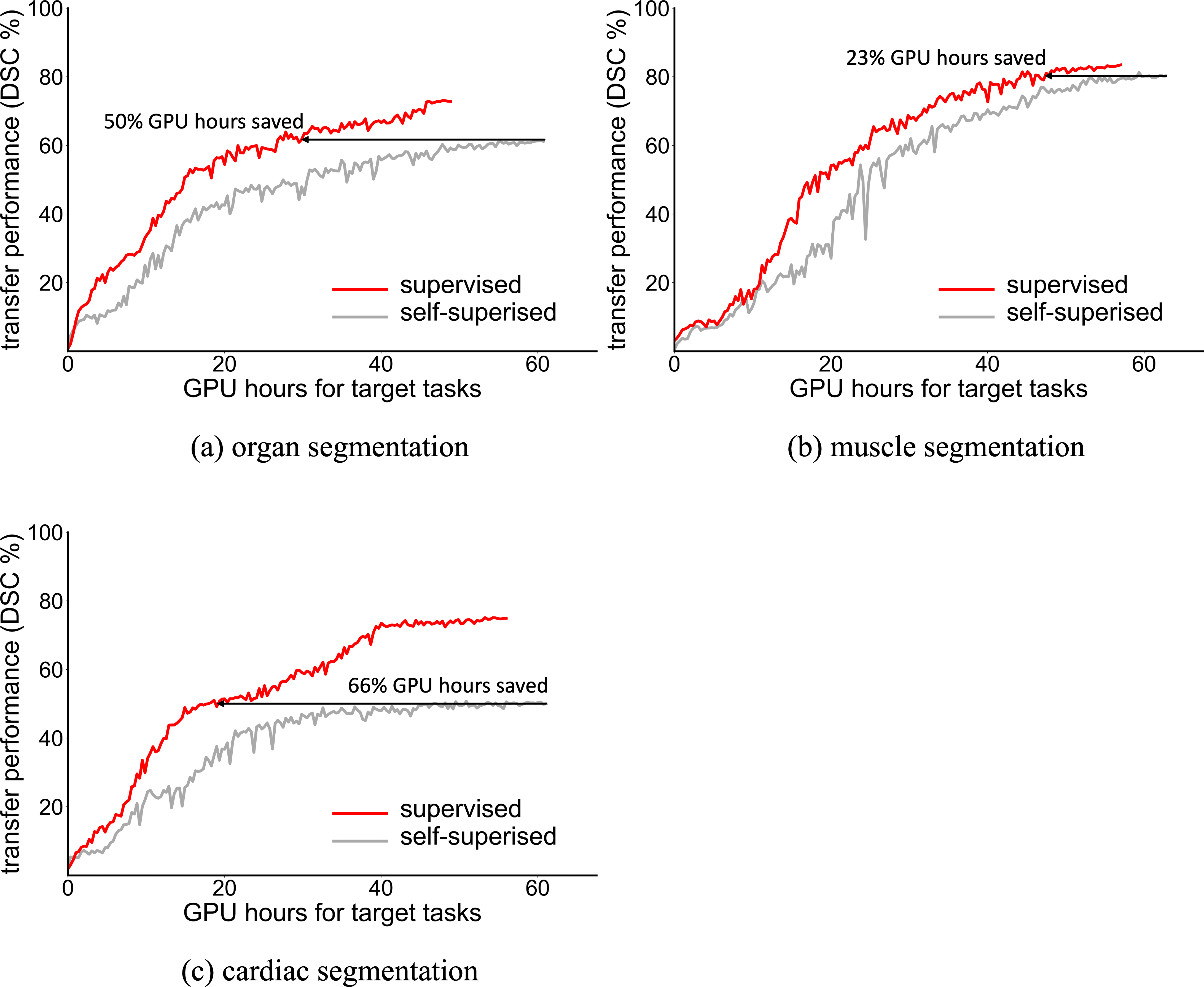}
    \caption{ 
    \textbf{Convergence \& learning efficiency in fine-tuning.} We present the learning curves for fine-tuning supervised and self-supervised models for three target tasks using 10\% of the training set. Supervised models achieve markedly better performance and converge faster than self-supervised counterparts by 50\%, 23\%, and 66\% for the tasks of organ, muscle, and cardiac segmentation, respectively.
    }
    \label{fig:computational_efficiency_appendix}
\end{figure*}

\clearpage

\section{Enhanced Features for Novel Datasets, Classes, and Tasks}\label{sec:feature_appendix}
\subsection{Direct Inference on Three External Datasets}\label{sec:direct_inference_appendix}

\begin{figure*}[h]
    \centering
    \includegraphics[width=1.0\linewidth]{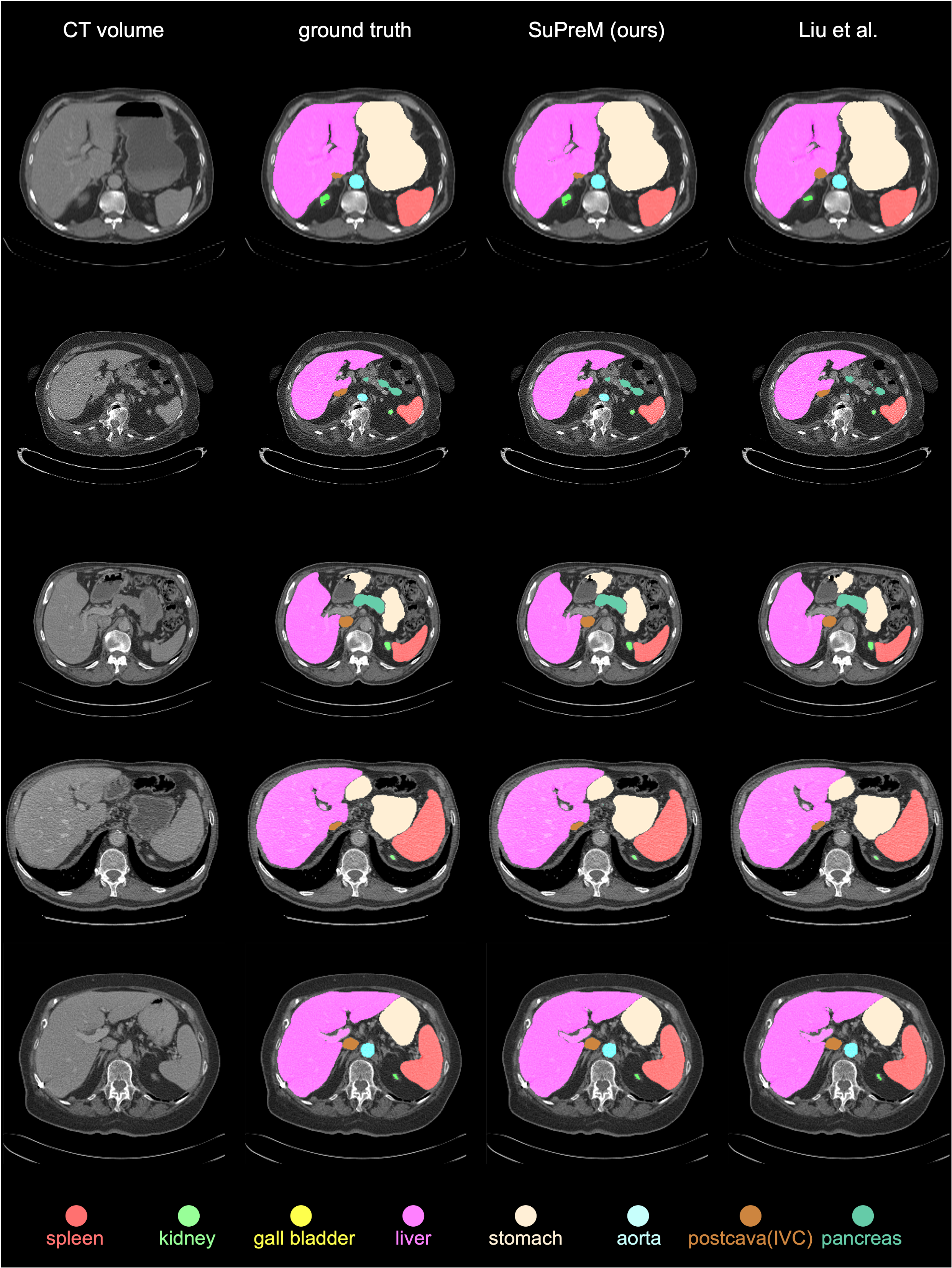}
    \caption{\textbf{Direct inference on TotalSegmentator.} We performed direct inference using TotalSegmentator, covering 104 classes. Here only 25 out of the 104 classes were visualized.
    }
    \label{fig:direct_inference_totalseg_appendix}
\end{figure*}

\clearpage
\begin{figure*}[h]
    \centering
    \includegraphics[width=1.0\linewidth]{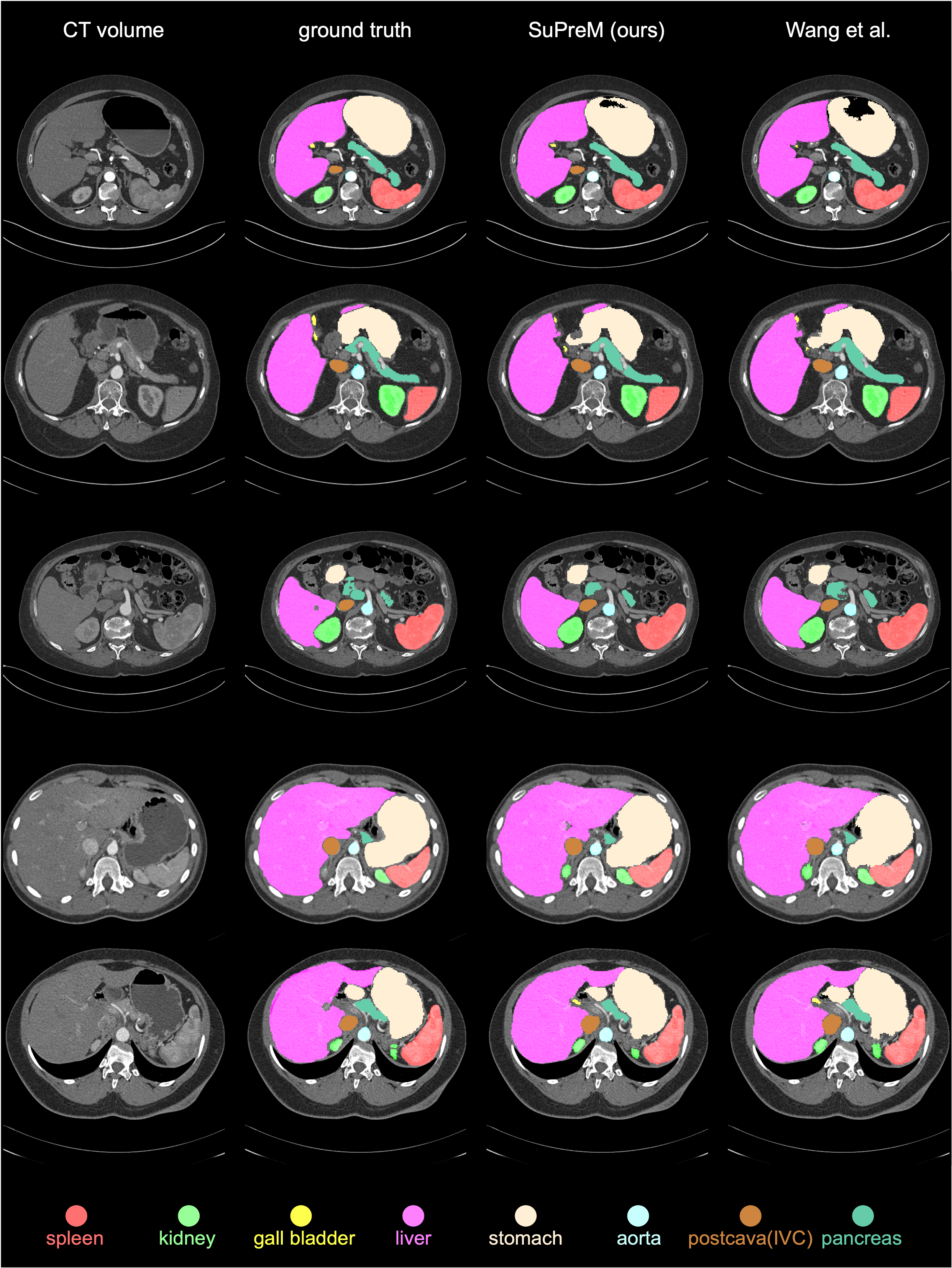}
    \caption{
    \textbf{Direct inference on the proprietary dataset.} We performed direct inference using a proprietary dataset at JHU, which covers 20 organ classes and three sub-types of pancreatic tumors. These include the aorta, adrenal gland, common bile duct, celiac abdominal aorta, colon, duodenum, gallbladder, postcava, left kidney, right kidney, liver, pancreas, pancreatic duct, superior mesenteric artery, small bowel, spleen, stomach, veins, left renal vein and right renal vein. The pancreatic tumor classes are pancreatic ductal adenocarcinoma (PDAC), pancreatic cysts, and pancreatic neuroendocrine tumors (PanNET). 
    }
    \label{fig:direct_inference_proprietary_dataset_appendix}
\end{figure*}

\clearpage
\subsection{Fine-tuning \ourmodel\ on Fine-grained Tumor Classification}\label{sec:novel_task}

\begin{figure*}[h]
    \centering
    \includegraphics[width=0.6\linewidth]{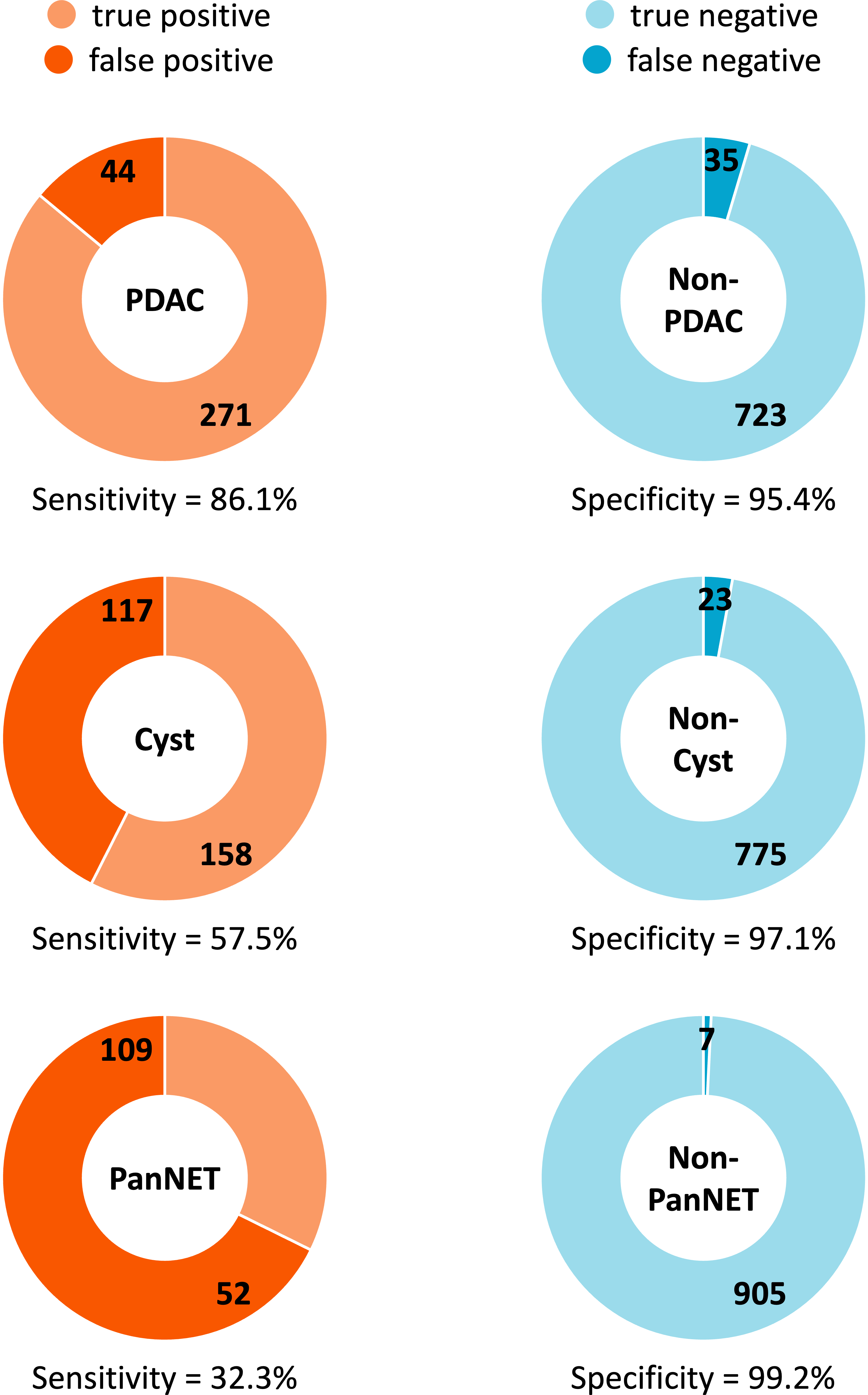}
    \caption{
    \textbf{Fine-grained pancreatic tumor classification.} We would like to stress the challenges in benchmarking tumor segmentation/classification, particularly due to the scarcity of annotations in publicly available datasets (often limited to hundreds of tumors). To overcome this limitation, we employed our proprietary dataset, which comprises 3,577 annotated pancreatic tumors, including detailed sub-types: 1,704 PDACs, 945 Cysts, and 928 PanNETs. The proprietary dataset contains CT scans taken by a variety of vendors, e.g., Philips, Siemens, GE, and Toshiba. This extensive dataset enabled us to thoroughly assess the transfer learning ability of our pre-trained models in tumor-related tasks. Notably, the transfer learning results detailed here demonstrate a sensitivity of 86.1\% and specificity of 95.4\% for PDAC detection. This performance surpasses the average radiologist's performance in PDAC identification by 27.6\% in sensitivity and 4.4\% in specificity, as reported in \citet{cao2023large}. This is one of the demonstrations of how our pre-trained models could be deployed for clinical applications.
    }
    \label{fig:tumor_classification_appendix}
\end{figure*}

\end{document}